\documentclass{emulateapj}
\shorttitle{Coronal lines and dust in SN~2005ip}
\shortauthors{Smith et al.}
\begin{document}

\title{Coronal lines and dust formation in SN~2005\lowercase{ip}: Not
 the brightest, \\ but the hottest Type II\lowercase{n} supernova}

\author{Nathan Smith, Jeffrey M.\ Silverman, Ryan Chornock, Alexei V.\
 Filippenko, Xiaofeng Wang, Weidong Li, Mohan
 Ganeshalingam, Ryan J.\ Foley\altaffilmark{1,2}, Jacob Rex, 
 \& Thea N.\ Steele}

\affil{Department of Astronomy, University of California, Berkeley, CA
 94720-3411; nathans@astro.berkeley.edu}

\altaffiltext{1}{Present address: Harvard-Smithsonian Center for
 Astrophysics, 60 Garden St., Cambridge, MA 02138.}
\altaffiltext{2}{Clay Fellow.}

\begin{abstract}

  We present optical photometry and spectroscopy of SN~2005ip for the
  first 3~yr after discovery, showing an underlying Type II-L
  supernova (SN) interacting with a steady wind to yield an unusual
  Type IIn spectrum.  For the first $\sim$160~d, it had a fast linear
  decline from a modest peak absolute magnitude of about $-$17.4
  (unfiltered), followed by a plateau at roughly $-$14.8 for more than
  2~yr.  Initially having a normal broad-lined spectrum superposed
  with sparse narrow lines from the photoionized circumstellar medium
  (CSM), it quickly developed signs of strong CSM interaction with a
  spectrum similar to that of SN~1988Z.  As the underlying SN~II-L
  faded, SN~2005ip exhibited a rich high-ionization spectrum with a
  dense forest of narrow coronal lines, unprecedented among SNe but
  reminiscent of some active galactic nuclei.  The line-profile
  evolution of SN 2005ip confirms that dust formation caused its
  recently reported infrared excess, but these lines reveal that it is
  the first SN to show clear evidence for dust in {\it both} the fast
  SN ejecta and the slower post-shock gas.  SN~2005ip's complex
  spectrum confirms the origin of the strange blue continuum in
  SN~2006jc, which also had post-shock dust formation.  We suggest
  that SN~2005ip's late-time plateau and coronal spectrum result from
  rejuvenated CSM interaction between a sustained fast shock and a
  clumpy stellar wind, where X-rays escape through the optically thin
  interclump regions to heat the pre-shock CSM to coronal
  temperatures.

\end{abstract}


\section{INTRODUCTION}

Core-collapse supernovae (SNe) show a variety of spectral properties
(see Filippenko 1997 for a review) based primarily on the amount of
mass shed by the progenitor, causing the outer layers of the star to
be stripped to different chemical layers at the time of its explosion.
In the Type~IIn subclass, substantial mass stripping has occurred
recently, leaving dense hydrogen gas in the circumstellar medium (CSM)
into which the SN blast wave propagates.  In this interaction, the
blast wave is decelerated and the CSM is illuminated, giving rise to
relatively narrow emission lines from the post-shock gas and the
unshocked CSM.

The Type~IIn subclass shows particularly wide diversity in both
luminosity and spectral features, depending on density, speed, and how
long before explosion the H-rich material was shed by the star.
SNe~IIn can be among the most luminous SNe observed if the CSM is very
dense, as in the case of SN~2006tf (Smith et al.\ 2008b) --- but they
can also be among the faintest SNe observed as in the case of the
so-called ``supernova impostors'' and related objects, which are not
yet clearly understood (e.g., Van Dyk et al.\ 2005; Thompson et al.\
2008).  In order for the CSM interaction luminosity to compete with
the main peak of the SN (arising from the diffusion of radioactive
decay luminosity and shock-deposited energy in the SN ejecta), the CSM
must be dense.  At lower densities where the conversion of shock
energy into visual light is less efficient, signs of weaker CSM
interaction might become more prominent in the spectrum once the
underlying SN fades; indeed, here we suggest that SN~2005ip was an
example of the latter.

SN~2005ip was discovered (Boles 2005) on 2005 Nov.\ 5.163 (UT
dates are used throughout this paper), located
2$\farcs$8~E and 14$\farcs$2~N of the center its host Scd galaxy
NGC~2906 ($\sim$2.1 kpc in projection on the sky).  Fox et al.\ (2008)
show images of SN~2005ip in its host galaxy.  With an apparent
redshift of $z = 0.00714$ (de Vaucouleurs et al.\ 1991), NGC~2906 is
located at a distance of roughly 29.7 Mpc (adopting $H_0 = 72$ km
s$^{-1}$ Mpc$^{-1}$).  Modjaz et al.\ (2005) reported that in a
spectrum obtained $\sim$1~d after discovery, SN~2005ip appeared to
be a normal Type~II event a few weeks after explosion, with a blue
continuum and broad absorption features indicating a SN-ejecta
expansion speed of roughly 15,400 km s$^{-1}$.  Recently, Fox et al.\
(2008) have shown that SN~2005ip had near-infrared (IR) excess
emission from hot dust, and suggested that grains formed in the
post-shock region, analogous to SN~2006jc (Smith et al.\ 2008a).

We present spectra showing that SN~2005ip also had narrow H$\alpha$
emission indicative of a Type~IIn classification, even on day 1, plus
an unusually rich forest of narrow coronal emission lines that
dominate the spectrum at later times.  Immler \& Pooley (2007)
reported a high X-ray luminosity at late times, implying a level of
CSM interaction comparable to that of other SNe~IIn like SN 1988Z.  We
present a brief discussion of the remarkable light curve and focus on
the spectral evolution of SN~2005ip, interpreting it as the result of
rejuvenated late-time CSM interaction in a steady wind with relatively
poor efficiency in converting shock energy to visual light.  We
confirm the suggestion by Fox et al.\ (2008) that dust formed, but our
analysis of line profiles reveals that dust formed in both the
post-shock shell and the fast SN ejecta at different times.

\begin{figure}
\epsscale{0.99}
\plotone{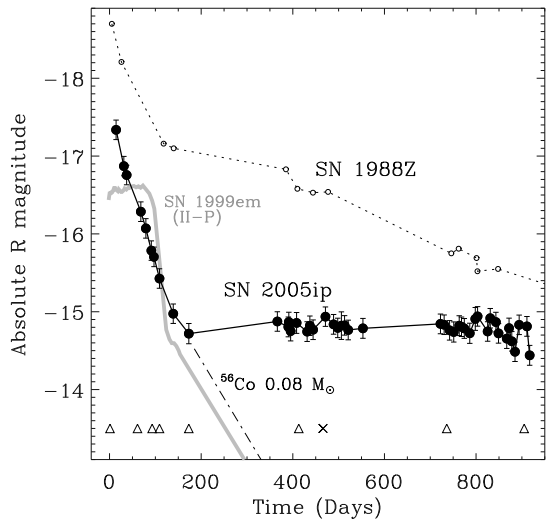}
\caption{The unfiltered (approximately $R$-band) absolute magnitude
  light curve of SN~2005ip measured by KAIT, compared to that of
  SN~1988Z (from Turatto et al.\ 1993).  The absolute magnitude of
  SN~2005ip is shown for a distance of 29.72 Mpc and an extinction of
  $A_R = 0.126$ mag (see text).  The small triangles at the bottom
  mark dates for which we secured spectra of SN~2005ip, and the ``X''
  marks the date (day 466) for which Immler \& Pooley (2007) observed
  its X-ray emission.  The gray curve shows a normal SN~II-P
  (SN~1999em) from our database for comparison, and the dot-dash line
  is $^{56}$Co decay luminosity for 0.08~M$_{\odot}$ of $^{56}$Ni.}
\label{fig:one}
\end{figure}

\begin{deluxetable}{lcc}\tabletypesize{\scriptsize}
\tablecaption{Photometry of SN 2005\lowercase{ip}}
\tablewidth{0pc}
\tablehead{
 \colhead{JD} &\colhead{mag.}  &\colhead{$\sigma$} }
\startdata
2453694.05  &15.15  &0.13  \\ 
2453711.03  &15.62  &0.13  \\
2453716.96  &15.73  &0.13  \\
2453747.91  &16.20  &0.13  \\
2453758.93  &16.42  &0.13  \\
2453770.91  &16.70  &0.13  \\
2453776.82  &16.79  &0.13  \\
2453788.84  &17.06  &0.13  \\
2453818.76  &17.52  &0.13  \\
2453852.70  &17.77  &0.13  \\
2454046.05  &17.62  &0.13  \\
2454070.04  &17.69  &0.13  \\
2454071.06  &17.62  &0.13  \\
2454074.97  &17.74  &0.13  \\
2454087.99  &17.63  &0.13  \\
2454110.93  &17.75  &0.13  \\
2454116.93  &17.67  &0.13  \\
2454123.88  &17.72  &0.13  \\
2454150.90  &17.55  &0.13  \\
2454168.80  &17.65  &0.13  \\
2454177.78  &17.70  &0.13  \\
2454185.71  &17.70  &0.18  \\
2454192.75  &17.67  &0.13  \\
2454200.74  &17.76  &0.13  \\
2454232.69  &17.70  &0.13  \\
2454402.02  &17.65  &0.13  \\
2454410.07  &17.66  &0.13  \\
2454422.07  &17.73  &0.13  \\
2454431.00  &17.75  &0.13  \\
2454443.04  &17.67  &0.13  \\
2454447.98  &17.68  &0.13  \\
2454454.06  &17.71  &0.13  \\
2454466.02  &17.77  &0.13  \\
2454478.00  &17.58  &0.15  \\
2454482.93  &17.55  &0.13  \\
2454504.89  &17.74  &0.13  \\
2454510.87  &17.57  &0.13  \\
2454522.91  &17.62  &0.13  \\
2454528.78  &17.77  &0.13  \\
2454546.79  &17.84  &0.13  \\
2454551.73  &17.70  &0.13  \\
2454557.77  &17.87  &0.13  \\
2454564.75  &18.01  &0.13  \\
2454573.72  &17.66  &0.13  \\
2454590.69  &17.68  &0.13  \\
2454596.68  &18.05  &0.13  \\
\enddata
\tablecomments{KAIT unfiltered photometry; roughly $R$ band.}
\end{deluxetable}

\begin{deluxetable}{lclccc}\tabletypesize{\scriptsize}
\tablecaption{Spectroscopy of SN~2005\lowercase{ip}}
\tablewidth{0pt}
\tablehead{
 \colhead{UT Date} &\colhead{Day\tablenotemark{a}} 
 &\colhead{Inst.\tablenotemark{b}}
 &\colhead{$\lambda$/$\Delta\lambda$}  &\colhead{Airmass} &\colhead{Exp.\ (s)} }
\startdata
2005 Nov. 06.65 &1    &DEIMOS &2500  &1.07 &100  \\
2006 Jan. 05.40 &61   &Kast   &600   &1.18 &1200 \\
2006 Feb. 06.40 &93   &Kast   &700   &1.19 &1800 \\
2006 Feb. 22.36 &109  &Kast   &600   &1.18 &1500 \\
2006 Apr. 27.26 &173  &LRIS   &1100  &1.03 &504  \\
2006 Dec. 23.56 &413  &DEIMOS &2500  &1.03 &500  \\
2007 Nov. 11.63 &736  &LRIS   &1500  &1.10 &600  \\
2008 Apr. 28.26 &905  &LRIS   &1500  &1.03 &600  \\
\enddata
\tablenotetext{a}{Days after discovery.}
\tablenotetext{b}{Kast, Lick 3-m; LRIS, Keck I; DEIMOS, Keck II.}
\end{deluxetable}

\section{OBSERVATIONS}

We monitored SN~2005ip photometrically using unfiltered images
obtained with the Katzman Automatic Imaging Telescope (KAIT;
Filippenko et al. 2001; Filippenko 2005) at Lick Observatory.  Its
host galaxy, NGC~2906, is included in the sample galaxies for the Lick
Observatory Supernova Search with KAIT.  As part of this search,
unfiltered images of NGC~2906 have been obtained in recent years, but
the field was behind the Sun before the discovery, so KAIT did not
yield a useful non-detection limit before explosion. These data also
constitute the follow-up photometry for SN 2005ip.

Flat fielding and bias subtraction were processed
automatically. Galaxy subtraction and differential photometry were
done using the KAIT pipeline (Ganeshalingam et al., in prep.),
although background host-galaxy emission is faint around SN~2005ip.
The unfiltered KAIT photometry is approximately equivalent to $R$-band
photometry (Li et al. 2003), and the resulting unfiltered magnitudes
are listed in Table~1.  The photometry is calibrated with the red
magnitudes of stars in the USNO B1 catalog (Monet et al.\ 2003), and
the uncertainty is the errors in photometry and calibration added in
quadrature (but dominated by the scatter in the calibration).
Adopting $m-M = 32.365$ mag and $A_R = 0.126$ mag for $E(B-V) = 0.047$
mag (Schlegel et al.\ 1998), the absolute $\sim R$-band magnitude
light curve is shown in Figure 1, where we take day zero to be JD =
2,453,679.66, the discovery date 2005~Nov.~5 (Boles 2005).

We also obtained visual-wavelength spectra of SN~2005ip on eight
separate dates using the Kast double spectrograph (Miller \& Stone
1993) on the 3-m Shane telescope at Lick Observatory, and using the
Low Resolution Imaging Spectrometer (LRIS; Oke et al.\ 1995) and the
Deep Imaging Multi-Object Spectrograph (DEIMOS; Faber et al.\ 2003) at
Keck Observatory (see Table 2). Except when the object was at low
airmass, the long slit (generally of width $\sim 2''$ at Lick and
$\sim 0.7$-$1''$ at Keck) was oriented along the parallactic angle
(Filippenko 1982) to avoid differential light losses due to
atmospheric dispersion.  Standard spectral data reduction was
performed for all epochs (e.g., Foley et al. 2003), including a
careful subtraction of background galaxy light by subtracting the
spectrum on either side of the SN.  Examples of the resulting spectra
are plotted in Figure 2 after having been corrected for $E(B-V) =
0.047$ mag.

There may be additional reddening and extinction local to SN~2005ip,
but the best way to correct for it is debatable.  The day 1 spectrum
shows local Na~{\sc i} absorption that may indicate additional
reddening of as much as $E(B-V) = 0.3$ mag (Munari \& Zwitter 1997),
or $A_R = 0.75$ mag assuming a standard value of $R = 3.1$.  However,
this narrow Na~I absorption feature is not clearly detected in later
spectra, and most importantly, is absent in the day 413 DEIMOS
spectrum with the same resolving power as on day 1 (although there is
an extra {\it emission} feature that develops on the red wing of
He~{\sc i} $\lambda$5876 on day 413, which complicates the
interpretation even further).  Since we see clear evidence for this
additional extinction only on day 1 and do not know precisely how
$A_R$ varies with time, we have not corrected the light curve in
Fig.~\ref{fig:one}.  If present, the way that this additional
reddening of as much as $E(B-V) = 0.3$ mag on day 1 would affect our
analysis below is as follows: (1) the peak absolute magnitude could
have been about 0.75 mag brighter than we quote below, (2) the
effective temperature we derive from a blackbody approximation of the
continuum on day 1 would be hotter, about 10$^4$ K instead of 7300~K,
(3) the initial photospheric radius would be smaller by about a factor
of 2, and (4) the corresponding constraints to the date of explosion
would be 4--5~d before day 1, rather than 8--10~d (\S 3.2).  None of
these corrections would alter our interpretation about the overall
nature of SN~2005ip or its progenitor.

\begin{figure*}
\epsscale{0.99}
\plotone{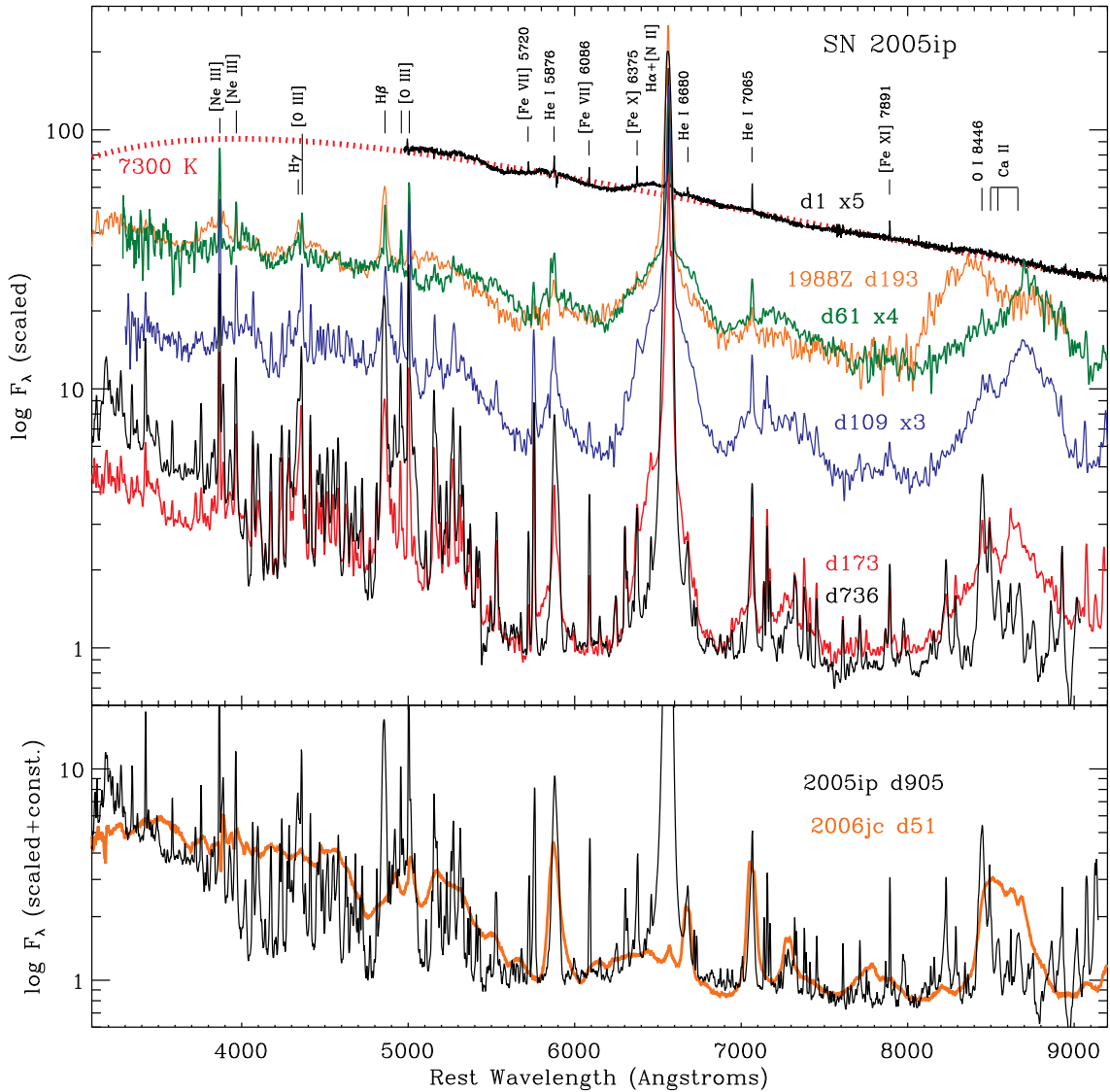}
\caption{{\it Top Panel}: Spectra of SN~2005ip on days 1 (black), 61
  (green), 109 (blue), 173 (red), and 763 (black).  For comparison, we
  also show the day 193 spectrum of SN~1988Z (orange) from our
  spectral database.  These have been corrected for reddening as noted
  in the text, and the days 1, 61, and 109 spectra have been scaled as
  noted in the figure for better visibility.  {\it Bottom Panel}: Same
  as above but showing the day 905 spectrum of SN~2005ip (black)
  compared to the day 51 spectrum of SN~2006jc (orange) from Smith et
  al.\ (2008a).  SN~2006jc is H-depleted and the lines are broader,
  but the unusual continuum shape is similar to
  SN~2005ip.}\label{fig:spec}
\end{figure*}

\begin{figure*}
\epsscale{0.99}
\plotone{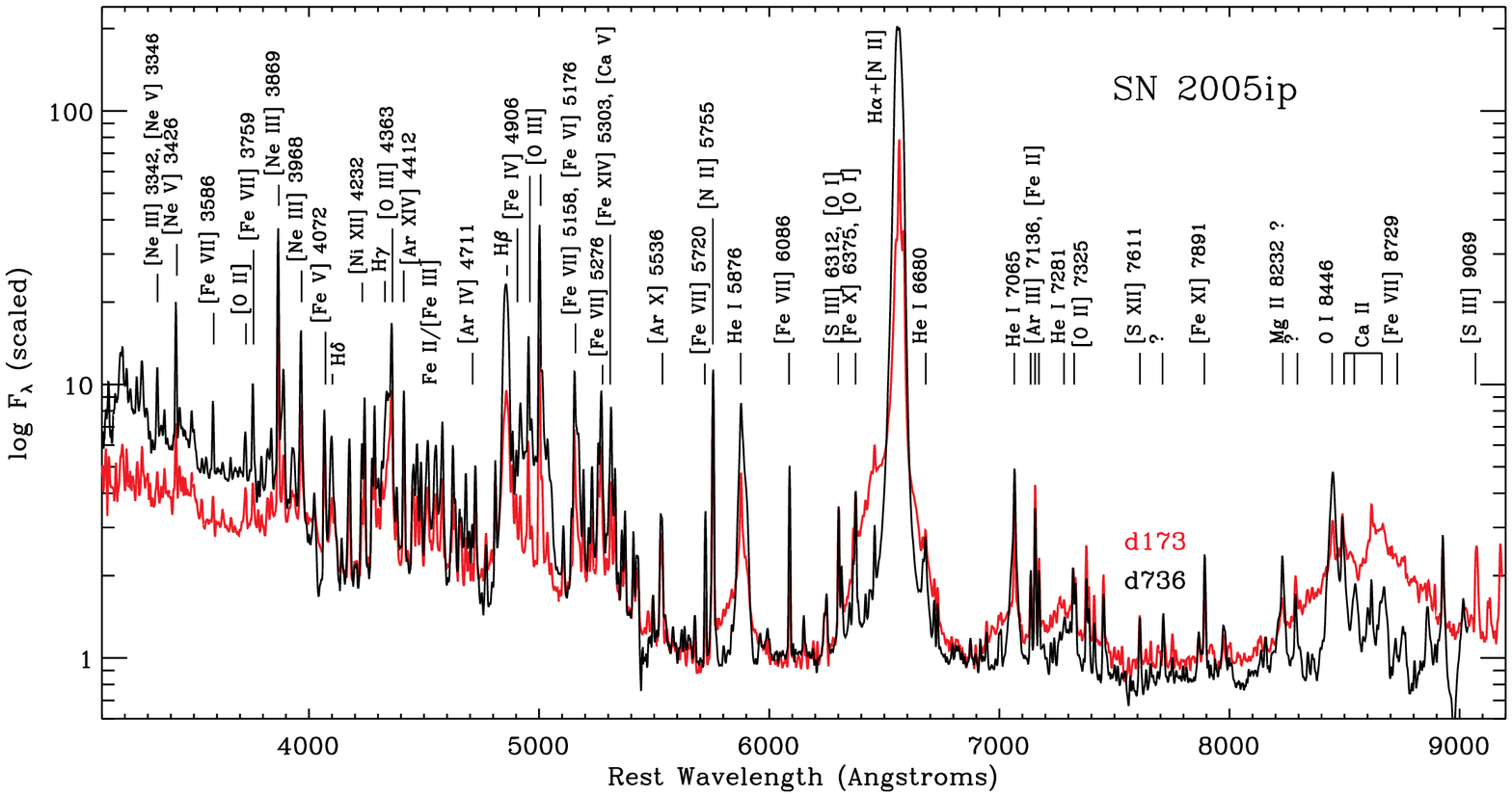}
\caption{Spectra of SN~2005ip on days 173 (red) and 763 (black), with
 a number of probable line identifications.}\label{fig:lines}
\end{figure*}

\section{RESULTS}
\subsection{The Unusual Light Curve of SN~2005ip}

The light curve of SN~2005ip (Fig.~\ref{fig:one}) shows a rapid
decline from its peak magnitude at the time of discovery, with the
early decline at $\sim$0.019 mag d$^{-1}$ reminiscent of SNe~II-L,
perhaps discovered a few weeks after explosion (in that case, its
peak magnitude may have been more luminous; e.g., Doggett \& Branch
1985; Arnett 1996).  If it was discovered near its peak, then its
luminosity was typical of average Type~II-P or II-L events
(Fig.~\ref{fig:one}), lacking the enhanced luminosity that is common
in strongly interacting SNe~IIn like SN 2006tf (Smith et al.\ 2008b) or
SN 1988Z (Turatto et al. 1993; Artexaga et al.\ 1999).  Attributing the
luminosity of the declining tail around day 150 to the radioactive
decay of $^{56}$Co would indicate a maximum initial $^{56}$Ni mass of
$\sim$0.08~M$_{\odot}$.  (The explosion date is not known, but if it
is, say, 3 weeks prior to discovery, then the implied $^{56}$Ni mass
could be $\sim$0.1~M$_{\odot}$.)

From $\sim$160 days after discovery until the present time, however,
the red light curve of SN~2005ip exhibits unusual behavior: its
decline hit a ``floor'' and remained essentially constant at an
absolute magnitude of roughly --14.8 (Fig.~\ref{fig:one}).  Slowly
declining SNe are rare, but to have a SN stay at constant visual
luminosity for several years after explosion is unprecedented.
SN~1988Z had qualitatively similar behavior with a similar kink in its
light curve, although SN~1988Z was more luminous than SN~2005ip and it
had a slow decline rather than a true plateau.  As we show below, this
late-time plateau of SN~2005ip is contributed in large part by strong
H$\alpha$ emission plus a number of other narrow lines over a flat and
weak red continuum.

\subsection{The Unusual Spectra of SN~2005ip}

Figure~\ref{fig:spec} shows the spectral evolution of SN~2005ip.  At
first glance, our earliest spectrum on day 1 is not very peculiar;
SN~2005ip has a normal Type II spectrum with a smooth blue continuum
and broad P~Cygni profiles for He~{\sc i} $\lambda$5876 and H$\alpha$,
indicating outflow speeds of almost 19,000 km s$^{-1}$ from the
blueshifted trough of H$\alpha$.  The continuum shape in the
dereddened spectrum on day 1 can be matched with a blackbody at 7300~K
($\pm$100~K; although see caveats in \S 2 regarding the effect of
possible additional reddening).  Its luminosity at that early time
indicates a characteristic blackbody radius of $1.4 \times 10^{15}$
cm.  This radius, combined with the observed expansion speed,
constrains the date of explosion to be at least 8--10~d before
discovery.

However, a closer look at the day 1 spectrum also reveals a sparse
series of narrow emission lines superposed on the spectrum (background
light from the host galaxy has been subtracted), so SN~2005ip was a
Type IIn event even at the earliest times.  The narrow lines observed
in our day 1 DEIMOS spectrum (5000--10,000 \AA) are H$\alpha$,
He~{\sc i} $\lambda\lambda$5876, 6680, 7065, [Fe~{\sc vii}] 
$\lambda\lambda$5720, 6086, [Fe~{\sc x}] $\lambda$6375,
and [Fe~{\sc xi}] $\lambda$7891 ([O~{\sc iii}] $\lambda$5007 may have
also been present but was near the blue edge of the detector where the
wavelength calibration is questionable).  These high-ionization lines
are commonly seen in active galactic nuclei (AGNs) (e.g., Filippenko
\& Sargent 1989), novae (e.g., Williams et al.\ 1991), and the solar 
corona (e.g., Wagner \& House 1968), but are rare in SNe (but see 
Gordon 1972).

Between days 1 and 150, SN~2005ip faded more rapidly than the
$^{56}$Co decay rate (Fig.~\ref{fig:one}).  During that time, it
developed a prominent broad H$\alpha$ line from the SN ejecta, with a
width near zero intensity of $\pm$16,000 km s$^{-1}$, while other
lines of intermediate width from the post-shock gas and very narrow
lines from the pre-shock CSM became more prominent.  The overall
character of the spectrum during this phase --- as well as the width
and strength of H$\alpha$ --- was very similar to that of SN~1988Z
(Filippenko 1991; Turatto et al.\ 1993; Stathakis \& Sadler 1991).  In
fact, Figure~\ref{fig:spec} shows that the only substantial difference
between the day 61 spectrum of SN~2005ip and the day 193 spectrum of
SN~1988Z is that the broad O~{\sc i} $\lambda$8446 feature from the
fast ejecta is stronger in SN~1988Z (the O~{\sc i} line strengthens
later in SN~2005ip).

By the time the underlying SN~II-L fades away (after day 160) and
SN~2005ip begins its long constant-luminosity phase, its spectrum has
undergone a remarkable transformation.  Narrow coronal lines now
dominate the appearance of the spectrum, along with some
intermediate-width lines from post-shock regions (He~{\sc i} and
Balmer lines).  Examples of the spectra dominated by these narrow
lines (days 173 and 736) are shown in Figure~\ref{fig:spec}.  These
coronal lines will be discussed in more detail below in \S 3.4.

\subsection{The Blue Pseudo-Continuum and SN~2006jc}

At early times, the shape of SN~2005ip's visual continuum can be
approximated by that of a $\sim$7300~K blackbody spectrum, typical of
SNe~II (Fig.~\ref{fig:spec}).  As the SN fades, however, narrow
emission lines begin to dominate the spectrum and the underlying
continuum takes on a different character.  At red wavelengths
($\lambda > 5500$~\AA), the continuum at later times is too flat to be
matched by any single blackbody shape, while shortward of 5500~\AA,
the continuum rises too steeply to be fit by any blackbody.  This
pronounced blue bump is apparent in the day 109 spectrum, it is well
developed by day 173, and it persists at least until our last
available spectrum on day 905 (Fig.~\ref{fig:spec}).

It is striking how well the continuum shape of SN~2005ip mimics the
blue pseudo-continuum and flat red continuum of SN~2006jc (Foley et
al.\ 2007; Smith et al.\ 2008a), and a comparison of the two is shown
in the bottom panel of Figure~\ref{fig:spec}.  With the exception of
H$\alpha$, the two match quite well (note that the strong and broad
Ca~{\sc ii} triplet in SN~2006jc comes from the underlying SN ejecta
and is present in earlier spectra of SN~2005ip).  This close match
implies two things, as follows.

First, if SN 2005ip were H-deficient and had higher speeds in the CSM
(broader lines), its spectrum might closely resemble that of SN~2006jc
(SN~2005la may be an intermediate case; Pastorello et al.\ 2008).
Since this blue feature dominates the visual broad-band luminosity of
SN~2005ip for $\sim$3 yr, it must arise as a result of luminosity
generated by ongoing CSM interaction and not radiation from the
underlying SN ejecta.  This was unclear in SN~2006jc, because the SN
faded more quickly (Foley et al.\ 2007).  Also, the fact that the blue
feature is seen in both SNe~2005ip and 2006jc means that it is not the
result of an increased burden of cooling on fluorescent lines caused
by the lack of H opacity, because SN~2005ip is not H-deficient.

Second, because the CSM lines in SN~2005ip are narrower, we see that
the blue pseudo-continuum can in fact be decomposed into a forest of
narrow emission lines.  Some regions of the spectrum with a
particularly dense clustering of narrow lines in SN~2005ip, such as
4000--4700 \AA, correspond to a bumpy blue ``continuum'' with an
overall flux distribution similar to that of SN~2006jc.  This overlap
therefore supports our earlier hypothesis (Foley et al.\ 2007; Smith
et al.\ 2008a) that the strong blue ``continuum'' in SN~2006jc was
fluorescence from a number of blended emission lines.  We emphasize
that the individual lines are seen in SN~2005ip only by virtue of
their formation in its slow 150--200 km s$^{-1}$ CSM, whereas the
progenitor of SN~2006jc was a Wolf-Rayet (WR) star with wind speeds of
order 2000 km s$^{-1}$.  With faster CSM speeds, the forest of narrow
lines in SN~2005ip would have blended together in a pseudo-continuum
as well.

The fact that the blue emission arises unambiguously in the CSM in
SN~2005ip has interesting implications for SN~2006jc.  Foley et al.\
(2007) first noted that in SN~2006jc, only He~{\sc i} lines that were
superposed on the blue pseudo-continuum had P~Cygni absorption
associated with them.  By analogy with SN~2005ip, one might conclude
that the P Cygni absorption arose in the cooler outer wind of
SN~2006jc's progenitor, while the blue pseudo continuum arises from
fluorescence in the inner wind or post-shock gas.  The lack of P Cygni
absorption then requires that the source of the red continuum was
cooler gas than that producing the He~{\sc i} lines, probably the
recombining SN ejecta.

A similar (although less pronounced) blue pseudo-continuum was also
seen in SN~1988Z (see Fig.~\ref{fig:spec}), as noted by Turatto et
al.\ (1993).  Turatto et al.\ pointed out the anomalously blue color
of SN~1988Z due to a peculiar rise at 3500--5700 \AA, as in SNe~2005ip
and 2006jc, and correctly speculated that it may be due to the overlap
of a huge number of emission lines.  Since both SNe~1988Z and 2005ip
showed coronal emission lines and the blue continuum, one might infer
a direct link between the two processes.  SN~2006jc, however, did not
exhibit the forbidden coronal lines for reasons that are unclear.

Another important point to note when comparing the continua of
SNe~2005ip and 2006jc is that while both showed the flat red
continuum, they behaved differently with time.  For a limited time,
SN~2006jc developed a strong red/IR excess indicative of hot dust at
$\sim$1600~K (Smith et al.\ 2008a); combined with the blue continuum,
this gave the visual-wavelength spectrum of SN~2006jc a ``U''-like
shape, which eventually disappeared.  SN~2005ip did not exhibit this
behavior, indicating that its dust was not hot enough or not luminous
enough to dominate the red continuum.  If the newly condensing grains
in SN~2005ip are significantly cooler than the $\sim$1600~K grains in
SN~2006jc, this may argue in favor of silicate grains in SN~2005ip
instead of the C-rich grains that probably formed in SN~2006jc (see
Smith et al.\ 2008a).

\begin{figure*}
\epsscale{0.99}
\plotone{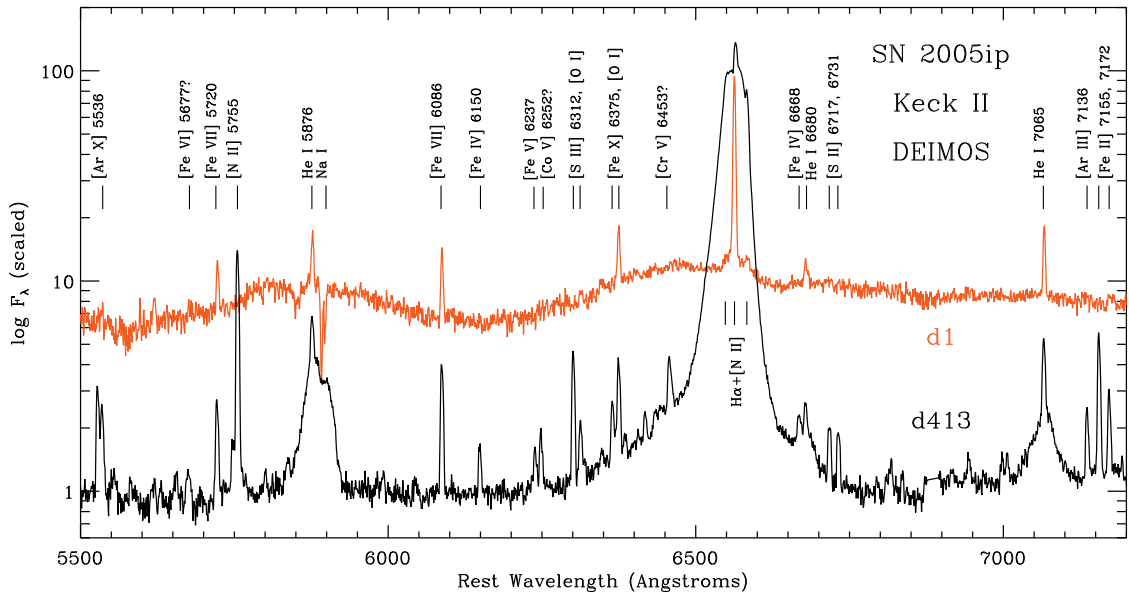}
\caption{A selected wavelength range around H$\alpha$ as seen in
 spectra obtained with high spectral resolution using the DEIMOS
 spectrograph on Keck II, on days 1 (gray; orange in the online
 edition) and 413 (black).  Coronal lines have only narrow
 components, whereas H$\alpha$ and He~{\sc i} also show
 intermediate-width components from post-shock gas at the later
 epoch, and broad H$\alpha$ emission is seen from the underlying fast
 SN ejecta.}\label{fig:deimos}
\end{figure*}

\subsection{Narrow Lines from the Highly Ionized CSM}

The many narrow lines that dominate the late-time spectrum of
SN~2005ip are noteworthy for their high ionization level and wide
range in ionization.  At late-times, many more of these
high-excitation lines have appeared, beyond the few visible on day 1
(Fig.~\ref{fig:spec}).  The narrow species seen prominently on day 1
were [Fe~{\sc vii}], [Fe~{\sc x}], [Fe~{\sc xi}], H$\alpha$, and
He~{\sc i}.  In addition to these, late-time spectra show [O~{\sc
  iii}], [Ne~{\sc iii}], [Ne~{\sc v}], [S~{\sc iii}], [S~{\sc xii}],
[Ca~{\sc v}], [Ar~{\sc iii}], [Ar~{\sc iv}], [Ar~{\sc x}], [Ar~{\sc
  xiv}], [Fe~{\sc iv}], [Fe~{\sc v}], [Fe~{\sc vi}], [Fe~{\sc xiv}],
[Ni~{\sc xii}], and others.  Identifications for several lines are
given in Figures~\ref{fig:lines} and \ref{fig:deimos}.

As noted earlier, these high-ionization lines are seen in novae, AGNs,
and the solar corona, but they are unusual in SNe.  In fact, no other
SN has exhibited such a rich spectrum of narrow coronal lines.  A few
of the stronger lines were seen in the spectrum of SN~1988Z (Turatto
et al.\ 1993), reinforcing the overall similarities between SNe~2005ip
and 1988Z that we have mentioned above.  However, the number of lines,
their strength, the range of ionization, and the highest-ionization
level seen are all higher in SN~2005ip.  Fransson et al.\ (2002)
presented a detailed analysis of narrow coronal lines in the SN~IIn
1995N, showing many of the same coronal lines we detect in SN~2005ip.
Again, though, the coronal lines in SN~1995N were not as prominent as
in our late-time spectra of SN~2005ip, and a comparison of the
evolution of coronal lines in the two SNe is not possible because it
is thought that SN~1995N was first discovered $\sim$10 months after
explosion.  In general, though, the physical conditions in SN~1995N's
CSM derived by Fransson et al.\ (2002) and the origin of its coronal
lines are similar to those from our analysis of SN~2005ip.

SN~1987A also showed faint emission from a subset of these coronal
features, including [Fe~{\sc xiv}] $\lambda$5303, in
intermediate-width lines from dense shock-heated gas (Fransson \&
Gr\"{o}ningsson 2007; Gr\"{o}ningsson et al.\ 2006, 2008).  This
emission appeared 10--20 yr after explosion, once the forward shock
began plowing into its dense ring nebula.  Some of these features may
also have been present in SN~1986J (Leibundgut et al.\ 1991), in the
earliest spectra of SN~1993J (Garnavich \& Ann 1994), and were seen as
weak narrow lines in the SN~IIn 1997eg (Hoffman et al.\ 2008).  While
these relatively weak coronal lines have been seen in a few other SNe,
the way that they dominate the spectral morphology of SN~2005ip is
unprecedented among known SNe.  Note that Figures~\ref{fig:spec} and
\ref{fig:lines} are plotted on a logarithmic scale, so the narrow
lines are quite strong compared to the continuum.

Emission from many lines with a wide range of ionization ---
especially the highest ionization --- implies high temperature gas
compared to that of most SNe.  The highest ionization species we
detect is the relatively weak [Fe~{\sc xiv}] $\lambda$5303 emission
line (at most epochs, it is severely blended with [Ca~{\sc v}]
$\lambda$5309).  The presence of [Fe~{\sc xiv}] $\lambda$5303 requires
that a substantial fraction of the CSM has electron temperatures up to
$2 \times 10^6$~K.  Most other lines suggest substantial amounts of
gas at several 10$^5$~K and below.  Since this coronal emission
dominates the overall spectrum to a greater degree than in any other
SN during its late CSM-interaction phase, we conclude that the bulk of
the emitting gas in SN~2005ip is hotter than in any other known
SN~IIn.

\subsection{Origin of the Coronal Lines}

One normally associates such wide ranges of ionization and temperature
with the post-shock cooling zone, but all of SN~2005ip's coronal lines
are very narrow, indicating that they cannot come from the post-shock
gas --- they must arise in the pre-shock CSM.  This is demonstrated
most clearly in Figure~\ref{fig:deimos}, which shows our two epochs of
high-resolution Keck/DEIMOS spectra on days 1 and 413 after discovery.
In Figure~\ref{fig:deimos}, one can see broad emission from the SN
ejecta in the broad component of H$\alpha$, intermediate-width
components emitted by post-shock gas in both H$\alpha$ and He~{\sc i}
lines, and only narrow emission components for all other lines.

The profiles of the narrow lines are unresolved in all of our Lick/Kast 
and Keck/LRIS spectra, but they appear to be resolved in our
high-resolution DEIMOS spectrum, with varying FWHMs of 120
(unresolved) to 240 km s$^{-1}$.  Lines this narrow are not formed in
the SN ejecta or in the post-shock gas like the intermediate-width
($\sim$1000--3000 km s$^{-1}$) lines usually seen in SNe~IIn (see,
e.g., Smith et al.\ 2008b; Chugai \& Danziger 1994).

Since many of the ions giving rise to the narrow lines have ionization
potentials corresponding to soft X-ray energies, and since their
narrow widths cannot be produced in the post-shock gas, we conclude
that the unusual spectrum of coronal lines in SN~2005ip arises from
pre-shock ionization of the CSM by X-rays emitted by the shocked gas.
Potential sources of X-rays are the initial shock breakout or
sustained X-ray emission from ongoing CSM interaction (or both), and
these will be considered in more detail below (\S 3.7) where we
discuss the observed line intensities.

\begin{figure}
\epsscale{0.98}
\plotone{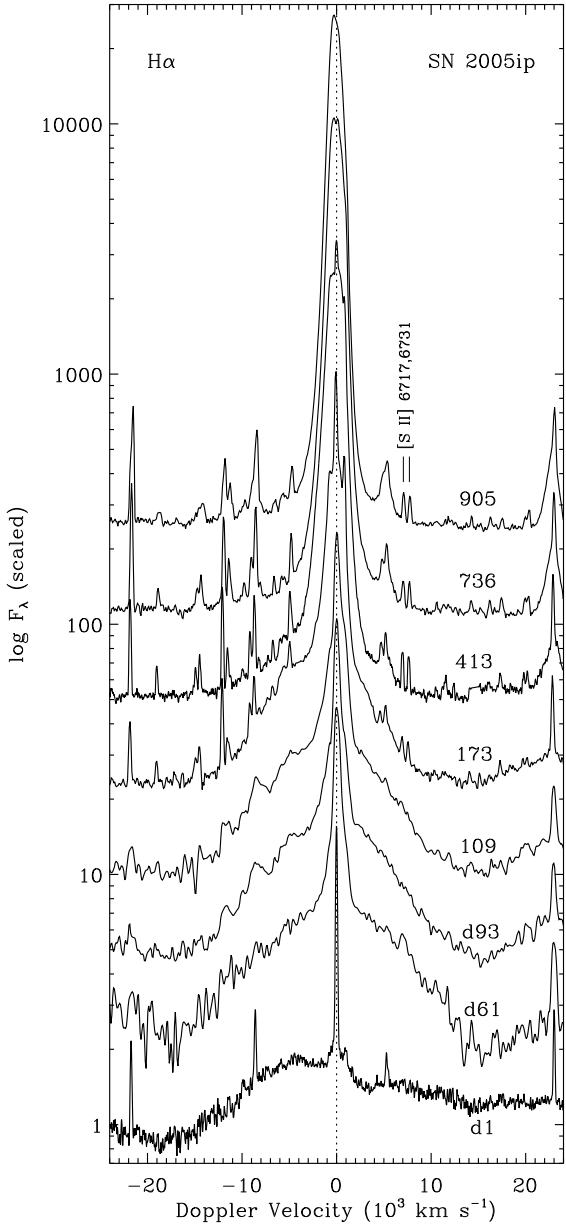}
\caption{H$\alpha$ line-profile evolution in SN~2005ip.  Changes can
  be seen in the relative strength of broad components ($\pm$16,000 km
  s$^{-1}$), as well as intermediate-width ($\sim$10$^3$ km s$^{-1}$)
  and narrow components.  Days 61--109 do not have sufficiently high
  dispersion to resolve the narrow components from the
  intermediate-width components.}\label{fig:profHa}
\end{figure}

\begin{figure}
\epsscale{0.98}
\plotone{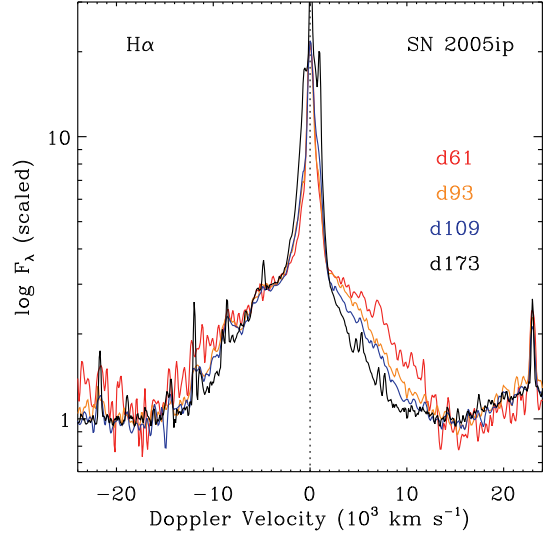}
\caption{Same as Fig.~\ref{fig:profHa}, but showing only days 61, 93,
  109, and 173 during the main luminosity decline of SN~2005ip, scaled
  and superposed on one another.  The red side of the broad H$\alpha$
  profile indicates dust formation within the expanding fast
  ejecta.}\label{fig:profHadust}
\end{figure}

\begin{figure}
\epsscale{0.92}
\plotone{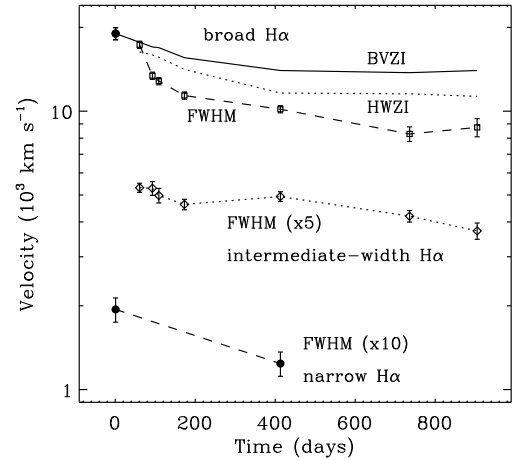}
\caption{Measured velocities for H$\alpha$ as a function of time. For
 the broad component (top), we show the velocity of the P Cygni
 absorption trough for day 1, followed by the blue velocity at zero
 intensity (BVZI), the half-width at zero intensity (HWZI), and the
 full-width at half maximum (FWHM) for the remaining epochs.  FWHM values 
 were measured with a Gaussian fit, but the fits were quite poor for the
 last two epochs because the broad component is much fainter than the
 intermediate-width component and is also crowded with several narrow
 coronal lines.  The FWHM of the intermediate-width component is
 multiplied by a factor of 5 for display here, and the FWHM of the
 narrow component (the narrow-line profiles are only resolved in our
 DEIMOS spectra on days 1 and 413) is multiplied by
 10.}\label{fig:velHa}
\end{figure}

\subsection{Line-Profile Evolution}

{\bf H$\alpha$ and H$\beta$:} The time evolution of the H$\alpha$
profile is shown in Figures~\ref{fig:profHa} and \ref{fig:profHadust}.
Ignoring the superposed coronal and other narrow lines, we find that
the underlying H$\alpha$ line profile shape, strength, and
evolution with time are remarkably similar to those of the H$\alpha$ 
line in SN~1988Z (Stathakis \& Sadler 1991; Turatto et al.\ 1993).

At most epochs, the H$\alpha$ profile of SN~2005ip can be separated
into three main components: (1) a broad component from the SN ejecta
(either unshocked or having just passed the reverse shock) within
typically $\pm$16,000 km s$^{-1}$, decreasing with time (see
Fig.~\ref{fig:velHa}), (2) an intermediate-width component at
$\pm$1100 km s$^{-1}$, and (3) a narrow component from the unshocked
CSM at $\pm$100--200 km s$^{-1}$.  In the day 1 spectrum, only
components (1) and (3) can be seen; the intermediate-width component
is absent.  The intermediate-width component steadily increases its
strength with time as the broad component fades, such that the
intermediate component comes to dominate the H$\alpha$ flux at late
times on the plateau of the light curve (note that
Fig.~\ref{fig:profHa} is logarithmic, so that at late times the broad
component is actually quite weak).  At late times the broad component
has faded and slowed such that it appears to merge with the wings of
the intermediate-width component and the two become difficult to
distinguish (electron scattering may also contribute to the line wings
here).  Also, at the latest epochs, the narrow CSM component of
H$\alpha$ disappears.

During the main luminosity decline of SN 2005ip in the first 170~d
after discovery, the broad component of H$\alpha$ develops an
increasingly asymmetric profile.  As shown in
Figure~\ref{fig:profHadust}, the red wing of the line fades more
quickly than its blue wing, producing a systematic blueward shift of
the line center.  This behavior can be attributed to dust grains that
increasingly and selectively block out the far side of the SN ejecta
as more dust forms.  Since we also note the possibility of dust
formation in the swept-up shell from He~{\sc i} line profiles (see
below; also see Fox et al.\ 2008), SN~2005ip may be the first clear
case where dust is seen to form in {\it both} the SN ejecta and the
swept-up post-shock shell.

\begin{deluxetable*}{lccccccccc}\tabletypesize{\scriptsize}
\tablecaption{H$\alpha$ Component Velocities, Total Flux, Equivalent Width, and Balmer Decrement}
\tablewidth{0pt}
\tablehead{
 \colhead{Epoch} &\colhead{Br.\ P Cyg} &\colhead{Br.\ BVZI}
 &\colhead{Br.\ HWZI} &\colhead{Br.\ FWHM}  &\colhead{Int.\ FWHM}
 &\colhead{Nar.\ FWHM} &\colhead{Flux} &\colhead{EW} &\colhead{H$\alpha$/H$\beta$} \\
 \colhead{} &\colhead{(km s$^{-1}$)} &\colhead{(km s$^{-1}$)} &\colhead{(km s$^{-1}$)}
 &\colhead{(km s$^{-1}$)}  &\colhead{(km s$^{-1}$)} &\colhead{(km s$^{-1}$)}
 &\colhead{(erg s$^{-1}$ cm$^{2}$)} &\colhead{(\AA)}  &\colhead{}}
\startdata
Day 1   &18500(900) &\nodata &\nodata &[16300]    &\nodata   &194(19) &1220$\times$10$^{-16}$ &49.5 &\nodata \\
Day 61  &17700(900) &17700   &16330   &17300(520) &1060(200) &\nodata &2810$\times$10$^{-16}$ &538  &19.1    \\
Day 93  &\nodata    &17000   &15920   &13400(400) &1060(300) &\nodata &2870$\times$10$^{-16}$ &1240 &10.6    \\
Day 109 &\nodata    &16900   &15620   &12800(380) &1000(300) &\nodata &2400$\times$10$^{-16}$ &1550 &10.8    \\
Day 173 &\nodata    &15600   &14130   &11400(340) &930(200)  &\nodata &1790$\times$10$^{-16}$ &3070 &9.8     \\
Day 413 &\nodata    &14000   &11620   &10200(300) &990(200)  &124(12) &2650$\times$10$^{-16}$ &5760 &11.6    \\
Day 736 &\nodata    &13800   &11550   &8300(500)  &840(200)  &\nodata &3250$\times$10$^{-16}$ &8880 &14.3    \\
Day 905 &\nodata    &14000   &11300   &8800(660)  &740(250)  &\nodata &3340$\times$10$^{-16}$ &9320 &24.7    \\
\enddata
\tablecomments{BVZI and HWZI uncertainties are roughly $\pm$500 km
 s$^{-1}$. The FWHM for the broad (Br.) emission component on day 1
 is in brackets because it is an underestimate of the true FWHM due
 to strong P Cygni absorption.  Uncertainties for the H$\alpha$ total
 flux and equivalent width are roughly $\pm$5\%.}
\end{deluxetable*}

\begin{figure}
\epsscale{0.92}
\plotone{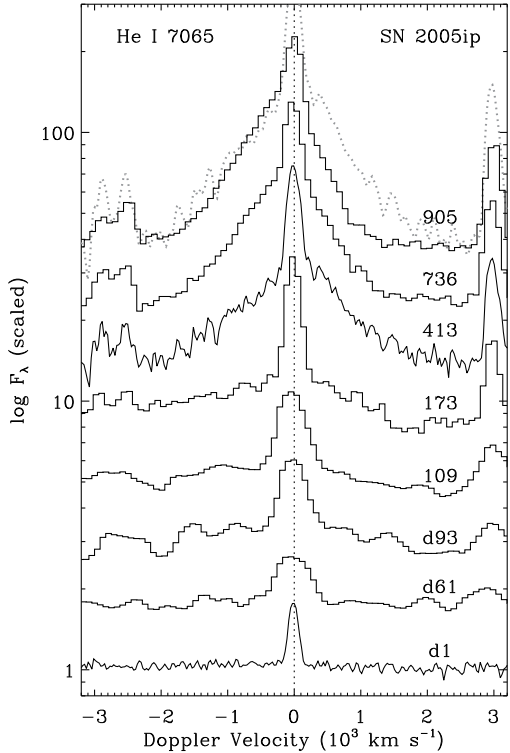}
\caption{Line-profile evolution of He~{\sc i} $\lambda$7065 in
  SN~2005ip.  Changes can be seen in the relative strength of the
  narrow component, as well as an increase with time in the
  intermediate-width (few 10$^3$ km s$^{-1}$) component.  Except for
  days 1 and 413, the narrow components are unresolved.  The gray
  dotted curve is the day 413 profile with a relatively symmetric
  intermediate-width component, plotted over the day 905 spectrum to
  demonstrate evidence suggesting new dust formation at late
  times.}\label{fig:profHe}
\end{figure}

\begin{figure}
\epsscale{0.92}
\plotone{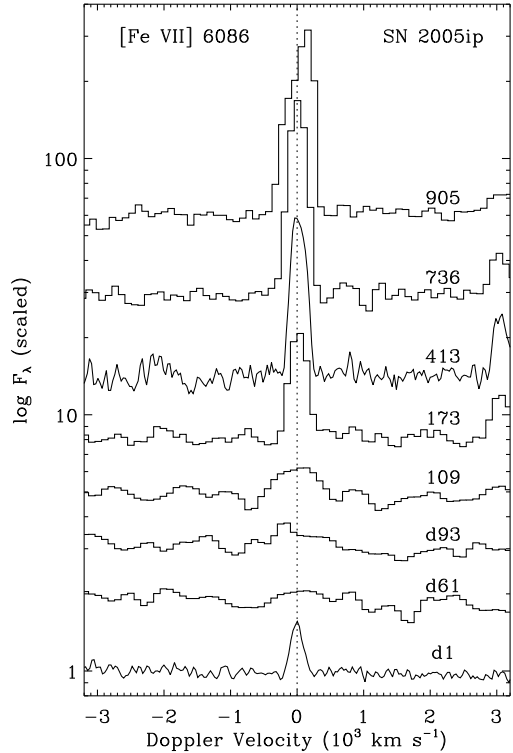}
\caption{Line-profile evolution of the coronal [Fe~{\sc vii}]
 $\lambda$6086 line in SN~2005ip.  Only the narrow component from the
 CSM is detected.  The only two epochs with sufficiently high dispersion 
 to resolve the narrow-component profiles are days 1 and 413, so the
 apparent increase in line width at late times may be a resolution
 effect.}\label{fig:profFe}
\end{figure}

The measured velocities for the three components of H$\alpha$ are
shown in Figure~\ref{fig:velHa} and listed in Table 3.  On day 1, the
broad component exhibits a smooth P Cygni profile; we show the speed
of the blueshifted absorption trough as a representative SN ejecta
velocity, but the intrinsic FWHM without blueshifted absorption could
be greater than this.  For the remaining epochs, no clear P Cygni
absorption feature is seen in the broad component.  Since the broad
line becomes asymmetric and is blended with the very strong
intermediate-width component, FWHM values derived from Gaussian fits
depend on the parameters one chooses for the fit.  As an independent
check, we also show the blue velocity at zero intensity (BVZI) and the
half-width at zero intensity (HWZI) for comparison.  The difference
between the BVZI and HWZI reflects the line's asymmetry, as noted
above.  One can see that the BVZI does not decline as fast or as much,
indicating that fast SN ejecta with speeds of $\sim$15,000 km s$^{-1}$
are still crossing the reverse shock at late times, even though the
FWHM suggests lower speeds.

The FWHM of the intermediate-width component from the post-shock CSM
gas drops from $\sim$1100 km s$^{-1}$ at early times to $\sim$800 km
s$^{-1}$ at late times.  Qualitatively, this deceleration of the cold
dense shell is expected from models of CSM interaction (e.g., Chugai
\& Danziger 1994).  The uncertainty in the intermediate-width
component's FWHM value is dominated by the unknown contribution of
[N~{\sc ii}] $\lambda$6583.  The uncertain contribution of [N~{\sc
 ii}] $\lambda$6583, in turn, makes it problematic to investigate
possible effects of post-shock dust formation on the red wing of the
intermediate-width component of H$\alpha$. To this end, we use 
He~{\sc i} lines instead (see below).

The narrow component also shows a reduction in the observed width with
time, from 190 km s$^{-1}$ on day 1 to 124 km s$^{-1}$ on day 413 (at
the same spectral resolution).  This may suggest that the progenitor's
wind speed increased in the decades immediately before the SN, or that
the luminosity of the SN itself accelerated CSM material at small
radii but had less influence on the more distant CSM.  This may also
explain why different narrow lines have different widths, if they
trace different radial zones of ionization and density in the wind.

{\bf He~{\sc i} lines:} We take the strong triplet line He~{\sc i}
$\lambda$7065 as representative of the behavior of He~{\sc i} line
profiles arising in the post-shock shell.  He~{\sc i} $\lambda$5876,
although strong, is problematic because its red wing is affected by
Na~{\sc i}~D.  The evolution of the He~{\sc i} $\lambda$7065 line
profile is shown in Figure~\ref{fig:profHe}. The relative strength of
the narrow component from the unshocked CSM stays nearly constant
during the main fading of the SN.  An intermediate-width component
from post-shock gas ($\pm$1800 km s$^{-1}$) is absent at early times
during the main light-curve peak; it first becomes conspicuous at day
173 and increases in its relative strength thereafter, coming to
dominate the total He~{\sc i} $\lambda$7065 flux at late times.  At
late epochs, this intermediate-width component becomes asymmetric,
with a deficit of emission in its red wing.  In
Figure~\ref{fig:profHe}, the asymmetric day 905 profile of He~{\sc i}
$\lambda$7065 is compared to the more symmetric profile on day 413
(dashed gray).  The relative fading of the red sides of the
intermediate-width He~{\sc i} line is reminiscent of that seen in
SN~2006jc (Smith et al.\ 2008a), and it implies dust formation in the
post-shock shell as we will discuss later in \S 4.1.

{\bf Coronal lines:} To investigate the evolution of line-profile
shapes for the high-ionization coronal lines, we consider the observed
behavior of [Fe~{\sc vii}] $\lambda$6086 (Fig.~\ref{fig:profFe}),
because it is relatively strong and not blended with other lines.  At
the earliest epoch (d1), the line profile is the same as for He~{\sc
  i} $\lambda$7065, indicating a pure line from the undisturbed CSM.
As time progresses, however, [Fe~{\sc vii}] $\lambda$6086 evolves very
differently from He~{\sc i} $\lambda$7065.  During the main luminosity
decline in the first $\sim$100~d, [Fe~{\sc vii}] $\lambda$6086 weakens
considerably with respect to the fading continuum.  It increases in
relative strength thereafter, but it never develops the
intermediate-width components that are seen in He~{\sc i} lines.
Instead, it remains as a narrow line emitted exclusively by the highly
ionized pre-shock CSM.  (Note that although the width of the line
appears to increase from day 413 to days 736 and 905 in
Fig.~\ref{fig:profFe}, this is a resolution effect attributed to the
lower dispersion of the days 736 and 905 spectra; see Table 2.)

\begin{figure}
\epsscale{0.98}
\plotone{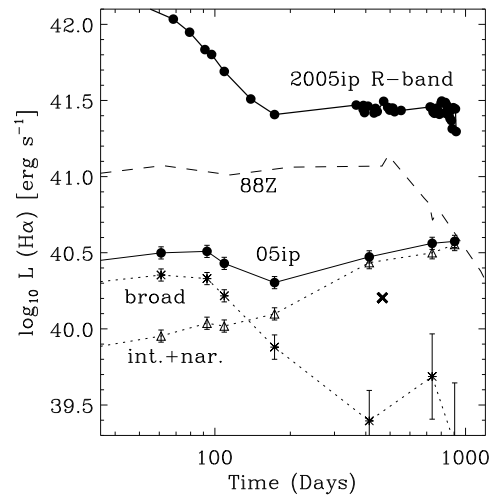}
\caption{The total H$\alpha$ line luminosity of SN~2005ip, compared to
  that of SN~1988Z (Artexaga et al.\ 1999).  We also show the
  bolometric luminosity of SN~2005ip derived from the $R$-band
  photometry, as well as the X-ray luminosity (shown by the ``X'')
  measured by Immler \& Pooley (2007).  Asterisks show the
  contribution of the broad component to the H$\alpha$ line of
  SN~2005ip, which decays with time, and triangles show the sum of the
  intermediate-width and narrow components, increasing with
  time.}\label{fig:haLUM}
\end{figure}

\begin{figure}
\epsscale{0.98}
\plotone{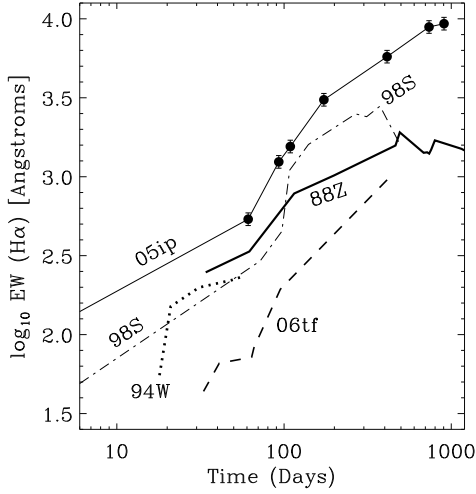}
\caption{The total H$\alpha$ equivalent width of SN~2005ip, compared
 to those of SNe~1988Z, 1998S, 1994W, and 2006tf (see Fig.\ 13 of
 Smith et al.\ 2008b and references therein).}\label{fig:haEW}
\end{figure}

\subsection{Line Intensities}

{\bf H$\alpha$:} Figure~\ref{fig:haLUM} shows how the H$\alpha$
luminosity of SN~2005ip changes with time, compared to its underlying
continuum and also with H$\alpha$ in SN~1988Z.  In both SNe, the total
H$\alpha$ luminosity remains roughly constant as the underlying SN
fades, although at late times SN~1988Z's H$\alpha$ luminosity begins
to drop, whereas that of SN~2005ip remains constant and even rises
slightly until the end of our observations.  Despite this nearly
constant luminosity, the line profile changes substantially
(Fig.~\ref{fig:profHa}).  Decomposing the total H$\alpha$ luminosity
into the broad component and the narrow plus intermediate-width
components (Fig.~\ref{fig:haLUM}) reveals two distinct phases. At
early times during the main fading of the underlying SN~II-L, the
broad H$\alpha$ dominates.  As it fades with the SN, the intermediate
and narrow components strengthen and come to dominate at late times.
At the latest epochs, however, Figure~\ref{fig:profHa} shows that the
narrow component disappears and H$\alpha$ is dominated by only the
intermediate component.

This reflects the fact that as the SN ejecta expand and cool, more H
atoms from the CSM are swept up into the post-shock shell and heated
by X-rays from that ongoing CSM interaction, probably mainly from the
reverse shock (Chevalier \& Fransson 1994).  Note that the {\it
  escaping} X-ray luminosity measured by Immler \& Pooley (2007) is
only half the H$\alpha$ luminosity at the same epoch
(Fig.~\ref{fig:haLUM}), suggesting that a large fraction (at least
half to 2/3) of the intrinsic X-ray luminosity is absorbed by the
swept-up shell, sustaining the optical luminosity, and indicating that
H$\alpha$ is an important coolant.

It is likely that the fraction of SN~2005ip's total X-ray luminosity
absorbed by the post-shock shell is lower than that of most other SNe~IIn
(i.e., it has a lower efficiency in converting shock kinetic energy
into light).  This is based on the fact that a substantial fraction of
the X-rays must escape the post-shock shell to reach the CSM and
energize the coronal spectrum that is more prominent than in any other
SN, and because SN~2005ip's observed X-ray luminosity is comparable to
that of SN~1988Z even though the visual CSM interaction luminosity is
less.

The lower efficiency of converting shock energy and X-rays into visual
light may be connected to the reason why SN~2005ip has a greater
H$\alpha$ equivalent width than other SNe~IIn (Fig.~\ref{fig:haEW}).
Specifically, a lower effective optical depth in the swept-up shell could
allow more X-rays to escape, and might also place a larger burden of
cooling on H$\alpha$ as opposed to the continuum.  That lower
effective optical depth might be due to either lower densities and a
lower progenitor mass-loss rate, or perhaps a higher degree of
clumping or asymmetry (see \S 4.2 and 4.3).

\begin{deluxetable*}{lcrrrrrrrr}
\tabletypesize{\scriptsize}
\tablecaption{Fluxes of Selected Narrow/Coronal Lines in SN~2005ip}
\tablewidth{0pc}
\tablehead{
\colhead{I.D.} &\colhead{$\lambda$} &\colhead{Flux d1} &\colhead{Flux d61} &\colhead{Flux d93} &\colhead{Flux d109} &\colhead{Flux d173}
               &\colhead{Flux d413} &\colhead{Flux d736} &\colhead{Flux d905} \\
\colhead{} &\colhead{(\AA)} &\colhead{(10$^{-16}$)} &\colhead{(10$^{-16}$)} &\colhead{(10$^{-16}$)} &\colhead{(10$^{-16}$)} &\colhead{(10$^{-16}$)}
           &\colhead{(10$^{-16}$)} &\colhead{(10$^{-16}$)} &\colhead{(10$^{-16}$)}}
\startdata

[Ne~{\sc iii}],[Ne~{\sc v}] &3342,46 &\nodata &\nodata 		&35.3(6.67)	&\nodata 	&(2.86) 	&\nodata        &15.3(1.81)	&13.6(4.21) 	\\
$[$Ne~{\sc v}]		&3426  	&\nodata       	&\nodata 	&16.1(6.64) 	&11.9(7.26)* 	&12.6(1.85)	&\nodata        &34.0(4.58)	&30.7(1.63) 	\\
$[$Fe~{\sc vii}]	&3586  	&\nodata       	&\nodata 	&(6.29) 	&(5.25) 	&5.10(1.50) 	&\nodata       	&10.3(1.44)	&8.45(1.07) 	\\
$[$O~{\sc ii}]		&3727  	&\nodata       	&45.5(12.1) 	&9.45(2.31) 	&8.35(2.31) 	&6.94(1.60)	&\nodata      	&7.34(1.20)	&5.69(0.94) 	\\
$[$Fe~{\sc vii}]	&3759  	&\nodata       	&(16.2) 	&12.9(4.37) 	&(5.06) 	&6.39(2.40)	&\nodata       	&16.1(1.48)	&12.4(1.85) 	\\
$[$Ne~{\sc iii}]	&3869  	&\nodata       	&153(13.9)	&105(4.81)	&112(7.22)	&60.0(5.26)	&\nodata       	&81.9(9.42)*	&62.2(5.84)* 	\\
$[$Ne~{\sc iii}]	&3968  	&\nodata       	&89.9(20.4)	&38.6(5.21)	&33.9(9.21)*	&30.7(7.03)*	&\nodata       	&29.0(1.89)*	&14.4(1.68)* 	\\
$[$Fe~{\sc v}]		&4072  	&\nodata       	&(17.3) 	&22.8(7.39)*	&22.8(1.65)*	&24.1(5.08)*	&\nodata      	&18.8(2.76)*	&11.7(2.15) 	\\
H$\delta$		&4103  	&\nodata       	&(1.29) 	&(5.07) 	&(4.49) 	&5.33(1.47)*	&\nodata       	&13.0(1.25)*	&19.7(3.21) 	\\
$[$Ni~{\sc xii}]	&4232  	&\nodata       	&(0.78)  	&(5.16) 	&6.36(2.41)*	&7.93(2.54)*	&\nodata       	&15.1(5.87)*	&11.0(3.99)* 	\\
H$\gamma$		&4340  	&\nodata       	&(1.74) 	&18.6(1.75) 	&17.6(2.47)*	&(3.10) 	&\nodata       	&52.8(14.3)* 	&37.0(8.93)*	\\
$[$O~{\sc iii}]		&4363	&\nodata       	&43.2(7.10)	&39.1(8.69)*	&39.7(4.79)*	&32.6(7.79)*	&\nodata        &33.9(4.77)*	&25.0(5.81)* 	\\
$[$Ar~{\sc xiv}]	&4412  	&\nodata       	&(1.08) 	&22.1(3.94)	&22.9(1.68)	&23.7(1.50)	&\nodata 	&21.5(0.81)	&12.6(1.06)	\\
$[$Ar~{\sc iv}]		&4711  	&\nodata       	&(0.85) 	&(10.9)	 	&(2.19) 	&(1.94) 	&(0.39) 	&2.98(0.65)* 	&1.72(0.76)*	\\
$[$Ne~{\sc iv}]		&4725	&\nodata	&10.9(2.11)	&(5.97)		&7.84(5.80)	&11.0(0.97)	&(0.36)	 	&11.7(0.67)*	&5.32(0.83)*	\\
H$\beta$		&4861  	&\nodata       	&66.7(10.8)	&26.8(3.81)*	&21.7(4.65)*	&11.5(2.20)*	&45.8(3.87)*	&190(15.8)*	&125(5.54) 	\\
$[$Fe~{\sc iv}]		&4906  	&(5.97)         &(6.11)		&(7.58)		&5.37(0.74)	&5.28(0.63)	&4.06(0.91)	&3.39(0.64)*	&2.87(1.51)*	\\
He~{\sc i}		&4922  	&(2.79) 	&(8.14) 	&(6.67) 	&7.70(4.84) 	&6.56(0.74) 	&4.49(0.69)	&7.64(0.99)*	&12.0(1.87)*	\\
$[$O~{\sc iii}]		&4959  	&(6.35)		&36.3(3.06)	&40.0(3.18)	&31.5(5.34)	&15.1(2.22)	&13.6(1.79)	&23.6(2.81)*	&14.6(3.23)*	\\
$[$O~{\sc iii}]		&5007 	&16.2(3.04)	&117(32.8)	&127(11.3)	&132(33.6)	&61.5(3.63)*	&43.6(4.68)	&79.7(11.1)*	&63.5(8.96)* 	\\
$[$Fe~{\sc vii}]	&5158   &(1.34) 	&12.3(5.84) 	&19.3(3.71)*	&15.8(4.76)*	&20.3(5.29)*	&14.9(1.96)*	&19.4(2.19)*	&13.4(3.17)* 	\\
$[$Fe~{\sc vi}]		&5176  	&3.23(1.57) 	&9.64(3.00)	&(5.09)		&(1.58)		&(1.94)	 	&(0.98)	 	&(2.84)	 	&3.10(0.85)*	\\
$[$Fe~{\sc vii}] 	&5276  	&(0.76) 	&19.6(7.47)* 	&23.3(2.40)*	&43.4(5.04)*	&25.6(2.65)*	&13.1(1.13)*	&29.0(3.69)*	&16.1(2.29)* 	\\
$[$Fe~{\sc xiv}],[Ca~{\sc v}]&5303,09&11.9(3.66)&7.70(4.85)	&13.6(4.39)*	&(1.77) 	&10.4(0.95)	&1.10(0.34)* 	&12.7(1.26)*	&8.59(1.84)* 	\\
$[$Fe~{\sc ii}] 	&5328   &(1.55)		&(7.94) 	&(2.09) 	&9.69(2.35)* 	&6.29(0.57)	&(0.29) *	&5.76(1.07)*	&2.63(0.52)*	\\
$[$Ar~{\sc x}]		&5536  	&(2.75) 	&(6.03) 	&7.06(0.85)* 	&\nodata*	&\nodata*	&4.72(0.71)*	&10.2(3.16)* 	&7.27(2.70)*	\\
$[$Fe~{\sc vii}] 	&5720  	&12.5(2.30)	&17.4(5.52)* 	&(3.33) 	&3.43(0.37)* 	&3.15(0.33)	&4.38(0.55)	&7.28(0.20)	&7.13(2.17) 	\\
$[$N~{\sc ii}]		&5755  	&(1.86)		&42.5(5.22)	&40.0(6.49)	&45.3(6.43)	&39.9(3.58)	&29.6(4.57)	&30.8(1.87)	&25.9(1.96)	\\
He~{\sc i} (Nar.)	&5876  	&17.3(2.29)	&83.9(7.66)*	&38.4(2.79)*	&17.9(2.22)*	&11.1(0.91)*	&8.01(0.87)*	&12.1(2.13)*	&15.8(1.25)* 	\\
He~{\sc i} (Tot)	&5876  	&17.3(2.29)	&83.9(7.66)*	&54.3(3.64)*    &49.8(5.34)*    &57.3(3.91)*    &76.24(3.56)*   &84.47(4.32)*   &91.0(2.8)*     \\
$[$Fe~{\sc vii}] 	&6086  	&18.1(2.54)	&10.3(1.80)* 	&9.2(1.22)* 	&8.29(1.50)*	&5.45(0.35)	&7.26(1.01)	&10.6(0.60)	&12.3(2.56) 	\\
$[$O~{\sc i}]		&6300  	&(1.71)		&(7.09)		&11.7(1.64)*	&6.27(1.06)*	&9.00(0.80)	&8.11(0.98)	&6.92(0.57)*	&5.83(1.59)*	\\
$[$S~{\sc iii}]		&6312  	&(3.41)		&(7.65)		&\nodata*	&7.07(0.54)*	&2.64(1.04)*	&1.99(0.26)	&3.18(0.78)*	&2.64(1.23)*	\\
$[$O~{\sc i}]		&6364  	&1.93(0.42)	&(7.07)		&\nodata*	&1.25(0.86)*	&3.49(1.56)*	&3.32(0.39)*	&2.99(1.47)*	&4.23(1.15)*	\\
$[$Fe~{\sc x}]		&6375	&24.6(2.98)	&6.13(1.59)*	&28.6(4.06)*	&11.6(1.22)*	&5.96(1.97)*	&6.83(1.13)*	&8.38(0.87)*	&7.14(1.26)*	\\
He~{\sc i}		&6680	&12.6(3.38)	&4.76(1.15)*	&(1.33)		&3.85(0.38)*	&2.44(1.49)*	&2.53(0.25)*	&4.53(0.49)*	&7.63(1.36)*	\\
$[$S~{\sc ii}]		&6717   &(2.19)		&12.1(1.95)	&(1.26)		&(1.18)		&1.82(0.08)	&2.22(0.55)	&1.62(0.22)*	&1.43(0.19)*	\\
$[$S~{\sc ii}]		&6731   &(1.33)		&(1.79)		&(1.30)		&(0.86)	 	&1.61(0.26)	&2.31(0.56)	&1.30(0.13)	&1.53(0.14)*	\\
He~{\sc i} (Nar.)	&7065	&24.6(1.93)	&41.3(4.53)	&24.7(2.15)	&20.2(1.65)	&10.2(0.72)	&7.92(0.72)	&6.73(0.34)	&6.43(1.04)	\\
He~{\sc i} (Tot.)	&7065	&24.6(1.93)	&41.3(4.53)	&39.1(2.63)     &35.6(1.69)     &21.3(0.91)     &29.8(1.44)     &29.9(2.67)     &30.6(2.24)     \\
$[$Ar~{\sc iii}] 	&7136  	&(1.64) 	&10.3(1.53) 	&6.11(0.92)*	&4.79(0.53)*	&3.75(0.64)	&3.08(0.55)	&2.89(0.27)	&2.62(0.29) 	\\
$[$Fe~{\sc ii}]		&7155  	&(1.95) 	&(2.65)		&9.25(0.95)*	&9.71(1.33)*	&12.5(0.80)	&10.4(1.61)	&6.91(0.56)	&5.67(0.65)	\\
He~{\sc i}		&7281  	&(2.50) 	&(2.92) 	&7.39(0.61)* 	&(1.69) 	&(1.57)		&1.17(0.29)*	&(0.41)	 	&2.54(0.66)*	\\
$[$O~{\sc ii}]		&7325  	&(1.88) 	&(4.23)		&13.5(3.49)*	&(2.14)		&(2.61)		&4.00(0.34)	&1.98(0.25)*	&1.89(0.15)*	\\
$[$S~{\sc xii}]		&7611  	&6.11(1.71)*	&(1.60) 	&5.12(0.38) 	&(2.03)		&2.11(0.29)	&2.67(0.69)*	&2.09(0.14)	&1.76(0.10)	\\
$[$Fe~{\sc iv}]?	&7704  	&(0.82) 	&(1.33) 	&3.66(0.40) 	&3.98(0.56) 	&2.97(1.01) 	&(0.29)	 	&(0.27)	 	&1.63(0.23) 	\\
$[$Fe~{\sc xi}]		&7891  	&13.4(1.34)	&(3.15)		&(1.37)		&3.61(0.33)	&4.73(0.73)	&5.12(0.74)	&4.63(0.60)	&4.80(0.23)	\\
He~{\sc i},[Fe~{\sc vi}]?&8232	&1.92(0.57)* 	&(0.98) 	&(1.80) 	&(1.01)		&2.86(1.00)	&2.89(0.48)	&7.48(2.05)*	&7.55(0.56)*	\\
He~{\sc i}        	&8295  	&(1.74) 	&(4.76) 	&(3.20) 	&2.51(0.20)* 	&3.68(0.65) 	&1.99(0.36)	&2.37(0.54)	&2.06(0.43) 	\\
O~{\sc i}		&8446   &2.01(0.58)	&26.7(3.14)	&16.3(2.31)*	&7.35(0.55)*	&8.75(1.67)	&9.78(2.24)	&29.8(1.90)*	&41.4(5.79)*	\\
$[$Fe~{\sc vii}] 	&8729  	&(1.66) 	&15.4(3.51)* 	&(2.75) 	&(0.64) 	&(0.62) 	&(0.39) 	&(0.37)  	&2.57(1.55)* 	\\
$[$S~{\sc iii}]		&9069  	&(2.14)		&22.8(1.51)	&14.9(1.76)	&10.5(1.22)	&11.3(2.12)	&4.84(1.07)*	&\nodata	&5.13(1.01)*	\\

\enddata
\tablecomments{Fluxes are given in units of 10$^{-16}$ erg s$^{-1}$
  cm$^{-2}$.  These are corrected for extinction and reddening of
  $E(B-V)$ = 0.047 mag in the Milky Way, but not for extinction local to
  SN~2005ip, which may be time dependent.  Uncertainties for measured
  line fluxes are in parentheses; an upper limit is in parentheses
  when no measurement is given. Listed uncertainties are 1$\sigma$
  based on the adjacent continuum noise, although the true
  uncertainties may be higher for blended lines, which are marked by
  *.}
\end{deluxetable*}

\begin{figure}
\epsscale{0.98}
\plotone{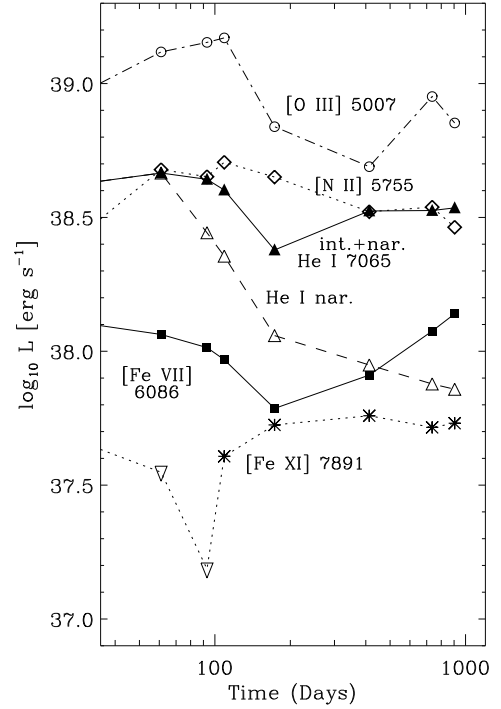}
\caption{Same as Fig.~\ref{fig:haLUM}, but comparing narrow-line
 luminosities of He~{\sc i} $\lambda$7065 (both narrow [unfilled
 triangles] and intermediate-width [filled]), [O~{\sc iii}]
 $\lambda$5007 (circles), [N~{\sc ii}] $\lambda$5755 (diamonds),
 [Fe~{\sc vii}] $\lambda$6086 (filled squares), and [Fe~{\sc xi}]
 $\lambda$7891 (asterisks). We do not plot [Fe~{\sc xiv}]
 $\lambda$5303 because it is blended with [Ca~{\sc v}]
 $\lambda$5309.}\label{fig:narrowLUM}
\end{figure}

{\bf He~{\sc i} and Narrow/Coronal Lines:} Figure~\ref{fig:narrowLUM}
shows the temporal behavior of luminosities for a few representative
narrow lines, while Table 4 lists the measured fluxes or upper limits
for many narrow lines.  A sparse sample of narrow coronal lines is
seen on day 1.  Some of these remain bright as the SN evolves, whereas
others initially fade to non-detectability and then gradually return
at late times.  Examples of the former are [Fe~{\sc vii}]
$\lambda$6086, [O~{\sc iii}] $\lambda$5007, and He~{\sc i}
$\lambda$7065, while examples of the latter behavior are [Fe~{\sc x}]
$\lambda$6375, [Fe~{\sc xi}] $\lambda$7891 and He~{\sc i}
$\lambda$6680.  Most narrow lines that appear later, with both
moderate and high ionization, have only upper limits on day 1 but then
strengthen dramatically as time proceeds.  This is particularly true
for lines of relatively low ionization (e.g., [O~{\sc i}], [O~{\sc
  ii}], [S~{\sc ii}]).

The variety of behavior seen in SN~2005ip's spectrum can be quite
complicated and perplexing.  Consider the evolution of a few lines in
particular: one sees very similar changes with time for the {\it
  total} flux in H$\alpha$ (Fig.~\ref{fig:haLUM}), He~{\sc i}
$\lambda$7065, [O~{\sc iii}] $\lambda$5007, and [Fe~{\sc vii}]
$\lambda$6086 (Fig.~\ref{fig:narrowLUM}). All four lines show an
initial ``hump'' in their light curves that has a broad peak just
before day 100, followed by marked decline following the SN continuum
luminosity to reach a minimum around day 200, and then all four lines
recover their strength at late times.  In H$\alpha$, the initial
``hump'' is emission from the {\it broad} component formed in the fast
SN ejecta, which then fades as the intermediate-width component takes
over the flux in the line (Fig.~\ref{fig:haLUM}).  For He~{\sc i},
however, the initial hump is dominated by the {\it narrow} component
which fades as the intermediate-width component comes to dominate
(Fig.~\ref{fig:narrowLUM}).  Both [O~{\sc iii}] and [Fe~{\sc vii}] are
seen exclusively as narrow lines at all times, never developing an
intermediate-width component even though their total intensity
recovers at late times like H$\alpha$ and He~{\sc i}.  It is
surprising that the total-flux evolution in all four lines is so
similar when their different subcomponents arise in three different
spatial zones within the SN --- the fast ejecta, the post-shock gas,
and the unshocked CSM, all with different ranges of density and
temperature.  We conjecture that the differing behavior in the
subcomponents of these lines reflects different ways of processing the
same energy source emerging mainly from the CSM interaction region,
but detailed radiative transfer models are probably needed for a
deeper understanding (e.g., Fransson et al.\ 2002).

In general, though, lines arising from a range of excitation levels
and a range of densities are seen at all times, with neutral atoms up
to high-ionization features like [Fe~{\sc xiv}].  One normally
associates such a wide range of ionization with post-shock cooling
zones, but as noted above, all these lines except H$\alpha$ and
He~{\sc i} are narrow lines emitted exclusively by as-yet unshocked
CSM gas.

From this qualitative fact we infer that the photoionized pre-shock
CSM of SN~2005ip must be clumpy or highly asymmetric, providing a
target with a range of densities and optical depths within the same
radial zone outside the forward shock.  The various coronal lines also
sample a wide range of critical densities up to 10$^7$--10$^8$
cm$^{-3}$, and this is presumably the reason why many lines are
suppressed in the day 1 spectrum --- in some cases transitions of the
same ionization levels that {\it are} seen on day~1.  For example, the
low ratio of [Fe~{\sc vii}] $\lambda$5159/$\lambda$6086 $\la$ 0.05 on
day 1 implies electron densities around 10$^8$ cm$^{-3}$ or higher
(Nussbaumer \& Storey 1982).  Later, during the plateau phase, various
[Fe~{\sc vii}] line ratios suggest similarly high densities and high
gas temperatures above 10$^5$ K.

Although a range of densities is present at all times, it is likely
that the average density falls at late times.  In that case, one would
infer that lines with lower critical densities that are collisionally
de-excited at early times would ``turn on'' only at late times, when
the shock and its photoionized precursor reach large radii where the
wind density is sufficiently low. The clearest example of this effect
is seen in the behavior of the [S~{\sc ii}] $\lambda\lambda$6717, 6731
and [O~{\sc ii}] $\lambda$3726, 3729 doublets (Fig.~\ref{fig:profHa}
and Table 4); these lines are absent until the start of the late-time
plateau phase when they both abruptly turn-on at day 173.  Both
transitions have critical densities of $n_c \approx 10^4$ cm$^{-3}$.
Much above that density, the lines are collisionally de-excited,
suppressing the line intensity roughly proportional to $n_c/n_e$.
Comparing the upper limits for [O~{\sc ii}] on day 109 to the flux
measured on day 173 then suggests minimum electron densities of
$\ga$3$\times$10$^5$ cm$^{-3}$ on day 109 and earlier.  After that
point, the [S~{\sc ii}] ratio reflects densities of a few 10$^2$
cm$^{-3}$.  It is likely that [S~{\sc ii}] and [O~{\sc ii}] trace
lower densities between clumps, while the narrow coronal lines sample
the densest regions in the clumps themselves.  The clumpy nature of
the progenitor wind is discussed further in \S 4.1.

In general, we can broadly categorize the observed evolution of line
intensities in SN~2005ip into three main phases:

(1) In the initial state on day 1, only a subset of narrow lines is
seen, probably corresponding to those with the highest critical
densities in the inner wind.  The absence of any low-ionization
features implies a very hard source of ionizing photons corresponding
to soft X-rays rather than to the far ultraviolet. There is not yet any
signature of strong CSM interaction in the day 1 spectrum (i.e., no
intermediate-width components of H$\alpha$ or He~{\sc i}), but these
features may be hard to see at early times due to the high luminosity
of the SN photosphere.  If the X-rays required to drive the coronal
lines were supplied by the X-rays arising from the initial shock
breakout (e.g., Matzner \& McKee 1999; Ensman \& Burrows 1992), this
would imply a progenitor radius much smaller than that of an extreme
red supergiant (RSG) as we infer later (\S 4.2).

(2) During the main decline of the SN light curve, lines either
disappear and recover slowly or have a ``hump'' where they strengthen
and then fade, as noted earlier.  These changes likely reflect a
combination transitional phenomenon, where the initial coronal
lines on day 1 respond to the different recombination timescales
appropriate to the range of ionization and densities in the clumpy
inner wind.  Simultaneously, as the forward shock expands, the
pre-shock CSM density drops and new features arise, responding to both
the dropping SN luminosity and the rising importance of luminosity
generated by CSM interaction.

(3) During the plateau phase, nearly all narrow forbidden lines
strengthen or remain steady (Fig.~\ref{fig:narrowLUM}).  This is true
for all narrow lines {\it except} permitted lines like He~{\sc i} and
H$\alpha$ that are also seen in the dense post-shock gas.  (Their
narrow components fade at late times.)  Sustaining the luminosity of
coronal lines with such high ionization potentials for $\sim$3 yr or
more, and in some cases causing them to get stronger, requires a
continued flux of soft X-rays.  Therefore, these lines cannot arise as
a result of photoionization by X-rays generated during shock breakout
alone.  Instead, the photoionizing X-rays to drive the late-time
coronal emission need to be generated in quasi-steady state by CSM
interaction.  An X-ray luminosity of $1.6 \times 10^{40}$ erg s$^{-1}$
was detected by Immler \& Pooley (2007) on day 466, consistent with
this hypothesis.

\section{DISCUSSION}

\subsection{The Progenitor's Mass-Loss Rate}

We estimate the mass-loss rate of the progenitor star using two
independent methods: the minimum mass in the CSM required to drain the
necessary energy from the SN, and the density of the CSM at a
particular radius based on emission lines from the pre-shock CSM.

In order for CSM interaction to tap the kinetic energy reservoir of
the expanding SN ejecta, the CSM must supply sufficient inertia.
Thus, the observed bolometric luminosity, if attributable to CSM
interaction alone, provides a minimum requirement for the mass-loss
rate of the progenitor star.  The required progenitor mass-loss rate
(e.g., Chugai \& Danziger 1994) is typically given by

\begin{displaymath}
\dot{M} = \frac{2 L}{\epsilon} \frac{V_w}{V_{SN}^3},
\end{displaymath}

\noindent where $\epsilon<$1 is the efficiency of converting shock
kinetic energy into visual light (an uncertain quantity), $V_w$ is
the progenitor's wind speed, $V_{SN}$ is the speed at which the CSM is
overtaken by the SN, and $L$ is the observed CSM-interaction
luminosity.  At late times after day $\sim$170, SN~2005ip shows a
constant-luminosity plateau at log$_{10}$($L$/L$_{\odot}$) = 41.45,
which we attribute to CSM interaction alone.  With a pre-shock speed
of $V_w \approx 120$ km s$^{-1}$, $V_{SN} = 10^4$ km s$^{-1}$, and
$\epsilon \la 0.5$ (an efficiency of 50\% is probably the most
optimistic value, although at lower optical depth the efficiency may
be less; see Smith \& McCray 2007), we have

\begin{displaymath}
\dot{M} = 2.1 \times 10^{-4} \big{(}\frac{\epsilon}{0.5}\big{)}^{-1} 
\big{(}\frac{V_w}{120}\big{)} \big{(}\frac{V_{SN}}{10^4}\big{)}^{-3} \
{\rm M}_{\odot} \ {\rm yr}^{-1}.
\end{displaymath}

An independent way to estimate the progenitor's mass-loss rate is to
infer values of the pre-shock wind density and speed at a specific
radius.  Earlier in \S 3.7 we estimated that the pre-shock wind
density illuminated by SN~2005ip on day 109 was
$\gtrsim$3$\times$10$^5$ cm$^{-3}$, based on the conjecture that the
narrow [O~{\sc ii}] and [S~{\sc ii}] doublets were suppressed until
that date (they appeared in our next epoch of spectra on day 173) due
to collisional de-excitation because of high densities in the
pre-shock CSM.  This minimum density represents the density of the
interclump wind, occupying most of the wind volume; the much higher
electron densities indicated by coronal lines correspond to the dense
clumps with a small filling factor.  On day 109, the BVZI of the broad
H$\alpha$ component was $\sim$16,900 km s$^{-1}$ (Table 3 and
Fig.~\ref{fig:velHa}).  This speed is seen in the fast SN ejecta near
the reverse shock, so the radius of the forward shock must be at least
$R \ga Vt = 1.6 \times 10^{16}$ cm.  At that radius, the inferred
density places another lower limit to the progenitor's mass-loss rate,
given by $\dot{M} = 4\pi R^2 n_e m_H V_w$, of roughly
$2.2 \times 10^{-4}$ M$_{\odot}$ yr$^{-1}$.

These two independent estimates give good agreement that the
progenitor star's mass-loss rate was of order $\dot{M} = 2 \times
10^{-4}$ M$_{\odot}$ yr$^{-1}$.  There are considerable uncertainties
involved, however.  The estimates above were really lower limits to
the required mass-loss rate because of efficiency and clumping,
respectively, so we therefore adopt $\dot{M}$ = (2--4) $\times
10^{-4}$ M$_{\odot}$ yr$^{-1}$ as a representative range of values for
discussion here.  This represents the progenitor's mass-loss rate for
at least 300~yr prior to core collapse.  Also, both estimates are
proportional to the inferred value of $V_w$.  We adopted $V_w \approx
120$ km s$^{-1}$ because this corresponds to the marginally resolved
width of the narrow-line components on day 413. As we noted earlier,
however, the faster velocities indicated at earlier times hint that
the pre-shock CSM could have been accelerated by the radiation force
of the luminosity from the SN itself (e.g., Chugai et al.\ 2002),
having a stronger effect at smaller radii probed at early times.
Whether the progenitor wind was relatively fast (120--150 km s$^{-1}$)
or slow (10--40 km s$^{-1}$) has important implications for the nature
of the progenitor star (see Smith et al.\ 2007), but doesn't change
many of the other conclusions we draw.

A more robust number is the wind density parameter $w = \dot{M}/V_w$
(making the derived result independent of $V_w$), which is equal to
roughly (1--2) $\times 10^{15}$ g cm$^{-1}$ for SN~2005ip.  By day 905,
the BVZI of H$\alpha$ still indicates speeds of $\sim$14,000 km
s$^{-1}$ seen in the vicinity of the reverse shock.  At that time, the
minimum radius is then $\sim 10^{17}$ cm.  With the values above, the
total CSM mass swept up by SN~2005ip on day 905 is at least
0.05--0.1~M$_{\odot}$ if the progenitor's mass-loss rate had been
constant.  SN~2005ip's CSM mass is considerably smaller than for many
other SNe~IIn, which have much higher values of $w\ga$10$^{16}$ g
cm$^{-1}$ and total masses that may be of order 0.5, 1, or even 10
M$_{\odot}$ (Chugai \& Danziger 1994; Smith et al.\ 2007, 2008b).
This seems intuitively reasonable for SN 2005ip, however, and paints a
self-consistent picture because SN~2005ip also had a relatively low
CSM-interaction luminosity compared to other SNe~IIn, and its low-mass
CSM apparently failed to significantly decelerate the bulk of the
fast SN ejecta.

\begin{figure}
\epsscale{0.98}
\plotone{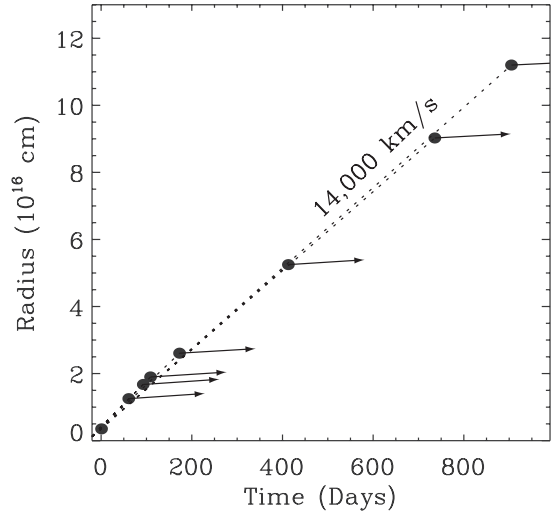}
\caption{Filled circles show the minimum radius of the reverse shock
  in SN~2005ip as a function of time (indicated by the observed BVZI
  of H$\alpha$), as well as the trajectories of the fast SN ejecta
  that are crossing the reverse shock at the observed epochs (dotted
  lines).  These are plotted assuming that day 1 is actually 21~d
  after explosion, although this value is not well constrained (see
  text).  The average expansion speed of the advancing reverse shock
  position is $\sim$14,000 km s$^{-1}$.  The solid arrows show the
  corresponding trajectories for slower gas giving rise to the
  intermediate-width components of emission lines.}\label{fig:radius}
\end{figure}

\subsection{The Progenitor's Clumpy Wind}

If the CSM fails to decelerate the forward shock, then how can one
explain the slower post-shock gas indicated by the persistent
intermediate-width components of H$\alpha$ and He~{\sc i} lines at
late times?  This problem is demonstrated in Figure~\ref{fig:radius}.
The intermediate-width lines are usually attributed to a cold dense
shell that forms at the contact discontinuity between the forward
shock and reverse shock (e.g., Chevalier 1982; Chugai 2001).  As time
proceeds, the fast SN ejecta that reach the reverse shock should have
progressively lower speeds appropriate for the radius of the cold
dense shell.  In SN~2005ip, the minimum radius where material crosses
the reverse shock advances at roughly 14,000 km s$^{-1}$
(Fig.~\ref{fig:radius}), which is much faster than the speed of the
cold dense shell at $\sim$1000 km s$^{-1}$.  This cannot be.

\begin{figure}
\epsscale{0.90}
\plotone{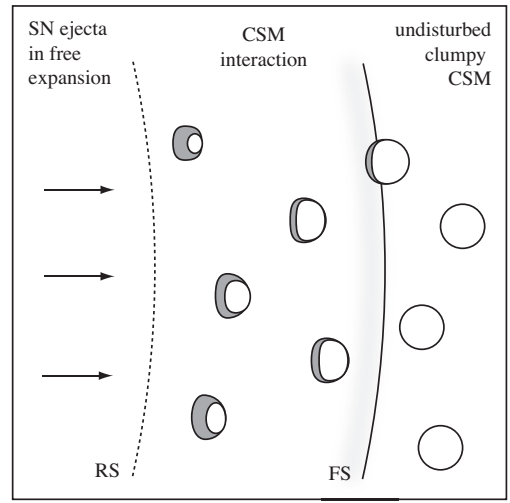}
\caption{A sketch showing a possible interpretation of the
  CSM-interaction region in a clumpy wind around SN~2005ip, analogous
  to the model proposed for SN~1988Z by Chugai \& Danziger (1994).  A
  fast forward shock propagates into the CSM in the less-dense regions
  between clumps.  As dense clumps in the CSM (white circles) are
  overtaken by the forward shock (solid curve), a slower shock is
  driven into each clump.  This interaction destroys the clumps in the
  time it takes the shock to cross the clump.  The dense post-shock
  gas in these clumps (gray) gives rise to the intermediate-width
  components of H$\alpha$ and He~{\sc i} lines.  X-rays from the more
  rarefied gas between the forward shock and reverse shock excite the
  narrow coronal lines in the unshocked regions of clumps on both
  sides of the forward shock.  The geometry of the reverse shock is
  unclear, because reflected shocks propagating back into the SN
  ejecta may complicate the geometry.}\label{fig:sketch}
\end{figure}

A possible solution to this inconsistency is that the progenitor's
wind is not homogeneous and spherical, but may be highly clumped or
asymmetric, as proposed for SN~1988Z by Chugai \& Danziger (1994).  An
example situation corresponding to a clumpy wind is sketched in
Figure~\ref{fig:sketch}.  In this scenario, there is no coherent cold
dense shell.  Instead, the forward shock maintains a high speed,
passing largely unhindered through the lower density regions between
clumps.  When the forward shock encounters a dense obstacle like a
clump or a dense equatorial disk, a much slower shock is driven into
the dense gas.  The clumps in the CSM may penetrate well into the
region engulfed by the expanding SN ejecta, surviving for a time
corresponding to the cloud crushing time $t_c \approx a/V_s$, where $a$
is the characteristic radius of a clump and $V_s$ is the speed of the
shock driven into the clump (about 1000 km s$^{-1}$).  With $a<<R$, we
have $t_c \approx 100$~d.  Thus, new generations of clumps in the
outer wind are likely to be continually overtaken by the forward shock
as older clumps are destroyed.  Since $V_s \approx 0.1 (V_{SN})$, the
clumps are likely to be about 10 times denser than the rarefied
interclump wind.

The post-shock gas in the clumps (gray in Fig.~\ref{fig:sketch}) is
denser still.  We attribute the intermediate-width components of
H$\alpha$ and He~{\sc i} to emission from this gas.  Extremely high
densities above 10$^9$ cm$^{-3}$ are indicated by the high
H$\alpha$/H$\beta$ ratio (Table 3) and the presence of these permitted
lines combined with the lack of any intermediate-width component in
any forbidden lines.  High densities lead to efficient cooling,
suppressing the coronal lines in the post-shock gas and potentially
allowing the gas to cool enough to form dust.  Similar densities
around 10$^{10}$ cm$^{-3}$ were inferred in the post-shock
dust-forming regions of SN~2006jc (Smith et al.\ 2008a).

Small-scale clumps embedded in an otherwise spherical wind with
mass-loss rates of the order required for SN~2005ip are reasonable and
expected from studies of evolved stars.  In fact, small-scale clumps
have now been directly detected in the CSM of the extreme RSG star
VY~CMa out to 10$^{17}$ cm around the star (Smith, Hinkle, \& Ryde
2009) --- this is the identical range of radii swept up by SN~2005ip.
Smith et al.\ (2009) also showed that the mass-loss rate of VY~CMa,
(2--4) $\times 10^{-4}$ M$_{\odot}$ yr$^{-1}$, and other conditions
are suitable to the production of a Type IIn event like SNe~2005ip or
1988Z.  Fransson et al.\ (2002) made a similar comparison to VY~CMa
for SN~1995N.  If this link to VY~CMa holds true, then it suggests
that SNe like 1988Z, 1995N, and 2005ip arise from fairly rare and
extreme RSGs that had initial masses of roughly 20--40 M$_{\odot}$.
Chugai \& Danziger (1994) have suggested that a slight depression in
the intermediate-width H$\alpha$ component may signal an equatorial
concentration as opposed to clumps; we see no sign of this effect
except for a very slight central depression on day 736.  Also, these
authors commented that the signature of large-scale clumps may be seen
in irregular structure of the intermediate-width component's profile.
We see no sign of this irregular structure either, consistent with our
claim that the clumps in question are an ensemble of small clumps,
distributed roughly symmetrically throughout the wind.

\subsection{Dust Formation in Both the Ejecta and Shell}

Recently, Fox et al.\ (2008) reported that SN~2005ip had a strong
near-IR excess over a time period of more than 900~d.  The $JHK$
fluxes were consistent with hot dust at 1400--1600~K.  Based on the
similar behavior of SN~2006jc (Smith et al.\ 2008a) and other
considerations, Fox et al.\ showed that the IR excess of SN~2005ip was
consistent with dust forming in the swept-up cold dense shell in the
post-shock layer, rather than an IR echo.

An unambiguous sign of the formation of new dust grains in a SN is
when one sees an increasing amount of extinction in the SN line
profiles.  Specifically, as an increasing amount of dust forms, it
should preferentially block more of the redshifted side of the ejecta.
This leads to a systematic fading of the red sides of emission lines
and a net blueshift of the line centroid.  So far, though, this effect
has only been clearly documented in a few cases, like SN~1987A
(Danziger et al.\ 1989; Lucy et al.\ 1989; Gehrz \& Ney 1989; Wooden
et al.\ 1993; Colgan et al.\ 1994; Wang et al.\ 1996; Moseley et al.\
1989; Dwek et al.\ 1992), SN~1999em (Elmhamdi et al.\ 2003), and
SN~2003gd (Sugerman et al.\ 2006; Meikle et al.\ 2007).  In all three,
the dust formed in the fast SN ejecta.

A second possibility is that in rare cases of SNe with strong CSM
interaction, dust grains may also form in the dense post-shock cooling
shell.  The strongest case for this type of dust formation was
SN~2006jc, which showed the characteristic blueshift in its
intermediate-width He~{\sc i} line profiles, and simultaneously
increased continuum extinction and IR excess emission from hot dust
(Smith et al.\ 2008a).\footnote{Although Nizawa et al.\ (2008)
 proposed that dust formed early in the ejecta of SN~2006jc, both
 Smith et al.\ (2008a) and Matilla et al.\ (2008) had found that dust
 forming in the fast SN ejecta could not account for the He~{\sc i}
 line-profile evolution, and concluded that the dust must form in the
 post-shock shell.}  Smith et al.\ speculated that this post-shock
dust formation occurred in SN~2006jc --- a peculiar Type Ib or Ibn ---
because the forward shock encountered a dense CSM shell ejected 2 yr
before, decelerating the shock and forming a dense ($\sim$10$^{10}$
cm$^{-3}$) post-shock layer that could cool fast enough to form dust.
Smith et al.\ conjectured that it may be easier to form this dust in
decelerated C-rich ejecta that have passed the reverse shock, although
this was difficult to prove from observations.  Post-shock dust
formation may also have been seen in the Type IIn events SN~1998S
(Pozzo et al.\ 2004) and SN~2006tf (Smith et al.\ 2008b).  In
hindsight, dust formation in the post-shock gas may also have caused
the fading of the red wings of intermediate-width H$\alpha$ components
in SN~1995N noted by Fransson et al.\ (2002).  This suggests that this
mode of dust formation may be more generic and may not require
C-enriched material.

As noted above in \S 3.6, SN~2005ip shows evidence for dust formation
in the post-shock shell based on the behavior of intermediate-width
He~{\sc i} line profiles (Fig.~\ref{fig:profHe}).  This, in principle,
confirms the recent suggestion by Fox et al.\ (2008).

There are two caveats we add, however.  First, suppression of the red
wings of the He~{\sc i} lines in SN~2005ip is only seen at late times,
more than 1~yr after explosion and later, whereas Fox et al.\ (2008)
reported a strong near-IR excess present much sooner after explosion
during the main decline in the light curve.  Thus, while we do see
evidence for dust formation in the post-shock gas at a time when an IR
excess is present, the IR excess was {\it already} present and did not
increase substantially when evidence for new dust developed in the
He~{\sc i} line profiles.  Therefore, we cannot confirm that the IR
excess before day 736 was due to post-shock dust formation, as opposed
to it having some other possible origin.  Second, as noted in \S 3.6,
we also see strong evidence for new dust formation occurring in the
fast SN ejecta based on the behavior of the broad H$\alpha$ profile
--- but it formed much sooner after the explosion, during the main
decline of the SN light curve.  Coincidentally, this is also when the
IR excess from hot dust was first seen by Fox et al.\ (2008).

Thus, from the line-profile evolution of SN~2005ip we conclude that
the IR excess from hot dust reported by Fox et al.\ (2008) can arise
from new dust grains in either the fast SN ejecta or the post-shock
shell, or perhaps a combination of both at different times.  There
seem to have been two episodes of dust formation: (1) dust formed
first in the rapidly expanding SN ejecta at days 60--170, causing the
initial near-IR excess seen at those times and affecting the broad
H$\alpha$ line profile, and (2) dust also formed much later in the
post-shock shell, sustaining the IR plateau at late times and shaping
the He~{\sc i} lines.  The luminosity source that re-heats the dust in
both zones may be generated by CSM interaction (propagating inward to
the dusty SN ejecta and outward to the cold dense post-shock gas).
This is consistent with the observation that the late-time visual and
IR luminosities are both $\sim$10$^{41.5}$ L$_{\odot}$.

SN~2005ip provides a unique case among known SNe in that it
shows strong evidence for dust formation in {\it both} the fast SN
ejecta and also in the slow and dense post-shock gas.  So far, the
SNe~IIn 2005ip, 1998S (Pozzo et al.\ 2004), and 2006tf (Smith et al.\
2008b), as well as the SN~Ibn 2006jc (Smith et al.\ 2008a), all seem
to have formed dust in their swept-up shells.  This hints that
post-shock dust formation may be common in SNe with strong CSM
interaction, and implies that we may need to revisit the cause of the
IR excess emission in SNe~IIn that has usually been interpreted as
light echoes, as noted also by Fox et al.\ (2008).  This difference is
important because dust that forms in fast SN ejecta or pre-existing
dust in the CSM is destined to be destroyed when it passes the reverse
shock or forward shock, respectively. Dust forming in the cold dense
shell, however, is {\it already} in the post-shock layer and will
likely survive.

In previous examples of SNe where dust formed in the post-shock gas,
it was assumed that this occurred in the cold dense shell between the
forward and reverse shocks, as in SN~2006jc (Smith et al.\ 2008a) and
SN~1998S (Pozzo et al.\ 2004).  In the clumpy model suggested here,
SN~2005ip does not have a coherent cold dense shell, so where does the
dust form in a scenario like that depicted in Figure~\ref{fig:sketch}?
In order to block the intermediate-width components at late times, the
dust may form in either the dense gas behind the shocks driven into
dense clumps or in the SN ejecta themselves.  As we noted earlier,
however, evidence for extinction from new dust is not seen in the
intermediate-width He~{\sc i} profiles until late times
(Fig.~\ref{fig:profHe}).  An intriguing possibility that may reconcile
the IR analysis by Fox et al. (2008) and the line-profile evolution
that we observe may be that dust forms in the dense post-shock gas
within individual clumps, but then gets ablated by and incorporated
into the expanding fast SN ejecta when a clump is eventually
destroyed.  This may provide an explanation for the dust present in
the SN ejecta that are otherwise presumably still too hot to form dust
at 60--170~d after explosion.  The radius we infer at days 60--170
when dust appeared is around 10$^{16}$ cm (Fig.~\ref{fig:radius}),
consistent with the minimum radius expected from the study of
SN~2005ip's IR emission by Fox et al.\ (2008).  A more detailed
analysis of this interaction is beyond the scope of this work, but it
is apparently consistent with the evidence we see for dust forming in
the fast SN ejecta as indicated by the evolution of the broad
H$\alpha$ profile.

This is also the first report of {\it both} post-shock dust formation
and strong coronal line emission in a SN, each linked to CSM
interaction, although Smith et al.\ (2008a) noted the similar
coincidence of dust formation and He~{\sc ii} $\lambda$4686 emission
in SN~2006jc and $\eta$ Carinae.  Classical novae also show signs of
dust formation and coronal emission, but these two are rarely seen
together in the same object (Gehrz et al.\ 1998).  The combination of
both dust formation and coronal emission has been reported only a few
times in classical novae (e.g., Gehrz et al.\ 1995, 2008; Mason et
al.\ 1996 and references therein).

\subsection{CSM Interaction in Context: A Comparison with SNe~1988Z,
 2006jc, and Others}

The most concise way to summarize the behavior of SN~2005ip is that it
is nearly a carbon copy of SN~1988Z, although less luminous (Stathakis
\& Sadler 1991; Turatto et al.\ 1993; Chugai \& Danziger 1994).  It is
likely that SN~1995N is closely related as well (Fransson et al.\
2002), but a comparison of the time evolution is difficult because
SN~1995N was discovered late.  The most remarkable similarities
between SNe~2005ip and 1988Z are their nearly identical H$\alpha$
profiles with three components (broad, intermediate-width, and
narrow), the way their velocities evolve with time, the parabolic
shape of the broad H$\alpha$ component, the lack of P Cygni absorption
(except in the broad component at day 1 in SN~2005ip), and the very
high (well above 10) H$\alpha$/H$\beta$ ratio.  Given these
similarities, we would predict that SN~2005ip should be an extremely
luminous radio source for years after explosion, akin to SN~1988Z (Van
Dyk et al.\ 1993; Williams et al. 2002) and SN~1995N (Chandra et al.\
2008).  The close comparison with SN~1988Z holds with the few
following caveats:

(1) SN~2005ip had a lower visual luminosity, both in its initial
decline and during its late-time luminosity plateau powered by CSM
interaction. At 1 yr after explosion it was about 2 mag
fainter, while after 2.5~yr this difference had reduced so that
SN~2005ip was only 0.5 mag fainter in visual light.

(2) Despite its lower visual luminosity, the X-ray luminosity of
SN~2005ip observed by Immler \& Pooley (2008) was comparable to that
of SN~1988Z, so SN~2005ip has a relatively high value of
$L_X$/$L_{Bol}$.

(3) The fastest speeds seen in the broad H$\alpha$ component were
similar in both SNe at early times, (19--20) $\times 10^3$ km s$^{-1}$,
but the speeds did not drop as much in SN~2005ip at late times.

(4) Although SN~1988Z did show some coronal lines, the coronal
spectrum in SN~2005ip was much more prominent and the coronal lines
were more numerous.  The coronal lines dominated the spectrum in way
that has not been seen in any previous SN, indicating pervasive hot
gas in the CSM.  The blue pseudo-continuum, composed of many blended
narrow lines, was also more prominent in SN~2005ip.

We suggest that all four of these differences compared to SN~1988Z can
be explained by the hypothesis that SN~2005ip's progenitor had a lower
mass-loss rate.  The wind density parameter we derive for SN~2005ip's
clumpy wind is about 5--10 times lower than that which Chugai \&
Danziger (1994) derived for SN~1988Z.  The lower wind density is less
able to decelerate the fast SN ejecta, explaining the sustained higher
speeds, and is therefore less able to drain kinetic energy from the
fast ejecta, resulting in a lower luminosity.  The lower density also
provides for a lower efficiency in converting X-rays to visual light
because of the lower optical depth within the post-shock layers.
Consequently, since it is less able to absorb the X-rays (and perhaps
also by producing a harder radiation field via the faster shock
speed), SN~2005ip generates a relatively stronger radiative X-ray
precursor that illuminates the pre-shock CSM and drives the coronal
spectrum.

We also noted some interesting similarities between SN~2005ip and
SN~2006jc.  The He~{\sc i} line profiles in SN~2006jc are similar to
those in the later stages of SN~2005ip.  Both objects also show
evidence for post-shock dust formation, plus a pronounced blue
continuum and flat red continuum that cannot be fit with any
blackbody.  By virtue of the narrow CSM lines in SN~2005ip, we see for
the first time that this blue pseudo-continuum (also present in
SN~1988Z to a lesser degree) is, in fact, composed of a large number
of blended emission lines originating in the CSM.  We had suspected
this in the case of SN~2006jc (Foley et al.\ 2007).  The blue
continuum is probably fluorescence of dense CSM gas illuminated by a
strong radiative precursor with a fairly hard spectrum, as in AGNs,
although a quantitative study of the origin of this emission remains
to be conducted.  Since SN~2005ip had H-rich CSM, this connection
proves that the similar blue continuum in SN~2006jc was not a result
of low opacities in H-depleted gas.

\subsection{Why Aren't There More SNe Like SN~2005ip?}

In many ways, SN~2005ip is what one might think that most SNe~IIn {\it
  should} look like.  The underlying SN was not extraordinarily
luminous, nor was the ongoing CSM interaction luminosity very high,
and the progenitor's clumped wind with a modest mass-loss rate of
order 10$^{-4}$ M$_{\odot}$ yr$^{-1}$ is not unusual among known
examples of luminous evolved massive stars.  The steady late-time
plateau suggests that this mass-loss rate was constant for centuries
before core collapse, consistent with the relatively steady winds seen
in the majority of massive stars.  Why, then, have we not seen a
SN~IIn quite like SN~2005ip before, with a steady late-time plateau
and such prominent coronal emission?

On the one hand, SN~2005ip had a relatively modest luminosity compared
with other more luminous SNe~IIn, and its late-time plateau was rather
dim at only $-$14.8 mag (or log$_{10}$[$L$/L$_{\odot}$] = 41.45).  It
was also relatively nearby at a distance of only $\sim$30 Mpc and its
host galaxy is faint, so it is reasonable to assume that there may be
more SNe~IIn with faint late-time plateaus like SN~2005ip that go
undetected at larger distances.  Many SNe~IIn do, however, fade more
than SN~2005ip in the few years after explosion (Li et al.\ 2002).

On the other hand, its peak luminosity did exceed that of a normal
SN~II-P, and a subset of the narrow coronal lines were in fact
prominent in the day 1 spectrum near maximum light.  Few SNe exhibit
these coronal lines, and in those that do, the coronal spectrum is
nowhere near as prominent as in SN~2005ip.  This suggests that the
origin of the coronal spectrum in SN~2005ip is the result of special
circumstances.

We speculate, therefore, that SN~2005ip sits near a critical
transition between the two regimes of normal SNe~II and SNe~IIn.  The
transition is defined by the progenitor star's mass-loss rate, such
that SN~2005ip probably marks the lower bound of wind density where
the wind is sufficiently dense to give rise to a Type IIn spectrum,
but not so dense that the resulting CSM interaction becomes very
optically thick.  Furthermore, the clumpy nature of the wind acts, in
effect, as a type of porosity that simultaneously provides dense gas
while allowing the fast blast wave and X-rays to escape through it
nearly unabated.  This seems to be an essential ingredient to
producing the coronal spectrum.

At higher progenitor mass-loss rates, the swept-up CSM would be more
optically thick, leading to higher efficiency in reprocessing shock
kinetic energy (and X-rays) into visual light as in SNe~2006tf,
2006gy, 2008es, and 2005ap (Smith et al.\ 2008b, 2007; Miller et al.\
2008; Quimby et al.\ 2007), but simultaneously preventing the escaping
X-ray flux that gives rise to the coronal spectrum in the pre-shock
CSM.  In the most luminous SNe~IIn with higher progenitor mass-loss
rates resulting from episodic pre-SN ejections (e.g., Smith \& Owocki
2006), the pre-SN mass loss is much higher than can be produced by a
normal stellar wind.  Thus, it is not clear that one expects those
more extreme CSM environments to have the same degree of clumpiness,
and the clumps might not survive in the optically thick post-shock
region where the shock kinetic energy is efficiently thermalized.

Altogether, given the moderately high progenitor mass-loss rate of
(2--4) $\times 10^{-4}$ M$_{\odot}$ yr$^{-1}$ in a clumpy wind,
consistent with an extreme RSG like VY~CMa (Smith et al.\ 2009), we
suggest that the progenitor of SN~2005ip was a moderately massive star
of roughly 20--40 M$_{\odot}$.  It probably exploded at the apex of
its RSG phase, when stars at that phase show the most vigorous mass
loss and the densest CSM environments (Davies et al.\ 2008).  In
context with other SNe~IIn, SN~2005ip's relatively modest luminosity
and presumed origin from a moderately massive RSG progenitor
underscore the much more extreme requirements for the luminous SNe~IIn
like SN 2006gy (Smith et al. 2007).  Namely, it seems consistent with
the idea that these even rarer and more extreme events arise from the
explosions of the most massive stars, requiring episodic or luminous
blue variable (LBV)-like mass ejections in their pre-SN phases (Smith
et al.\ 2007).  SN~2005ip's progenitor was probably also less massive
than that of the SN~IIn 2005gl, which Gal-Yam et al.\ (2007) found to
be consistent with a very massive LBV-like star.

\section{CONCLUSIONS}

We present and analyze visual-wavelength photometry and spectroscopy
of SN~2005ip obtained at the Lick and Keck Observatories during a
period of $\sim$3 yr after explosion.  Our main conclusions are as
follows.

(1) The data show that SN~2005ip was composed of an underlying
broad-lined Type II-L event that dominated the luminosity decline
during the first $\sim$160~d, superposed with a Type IIn spectrum
arising from constant-luminosity CSM interaction that caused a
remarkably steady late-time plateau.  SN~2005ip was not unusually
luminous compared with other SNe~IIn, and its initial mass of $^{56}$Ni
was less than about 0.1~M$_{\odot}$.

(2) As the underlying SN~II-L faded, the spectrum of SN~2005ip came to
exhibit a forest of narrow coronal emission lines, the number and
strength of which dominated the spectrum to an unprecedented degree.
Ionization levels as high as [Fe~{\sc xiv}] are seen.  These forbidden
coronal lines arise exclusively in the pre-shock CSM, showing no
broader components from the post-shock gas or SN ejecta.  Only a
subset of the narrow coronal lines was present in the day 1 spectrum.

(3)  Following the discovery by Fox et al.\ (2008) that SN~2005ip
showed near-IR excess emission attributable to freshly synthesized
dust, we confirm evidence for this dust formation from its influence
on the evolution of visual-wavelength emission lines.  Components of
both He~{\sc i} $\lambda$7065 and H$\alpha$ have red wings that fade
faster than the constant blue wings, as expected if new dust grains
preferentially block the far side of the object.  However, these two
tracers reveal dust at different places at different times as the SN
evolves.  At very late times (2--3 yr after explosion),
intermediate-width components of He~{\sc i} lines reveal dust forming
in a post-shock shell analogous to the post-shock dust formation in
SN~2006jc (Smith et al.\ 2008a), as proposed by Fox et al.\ (2008).
At early times, however, the situation is qualitatively different: we
see no evidence for dust in the intermediate-width components of these
lines, but instead, we see the same fading of the red wing in the 
{\it broad} component of H$\alpha$.  This means that at early times,
either (a) the dust formed directly in the fast SN ejecta, or (b) the
dust formed in the post-shock gas of individual clumps, but was
quickly incorporated into the rapidly expanding SN ejecta as the parent
clumps were ablated and destroyed.  At both epochs the dust in both
locations is heated primarily by the constant luminosity from CSM
interaction.

(4)  The photometric and spectroscopic evolution of SN~2005ip was most
similar to that of the strongly interacting and X-ray/radio-bright
SN~IIn 1988Z, especially with regard to the evolution of its H$\alpha$
profile.  The close comparison holds with the exceptions that
SN~2005ip was less luminous, had a higher $L_X$/$L_{Bol}$ ratio than 
SN~1988Z, and exhibited stronger narrow coronal lines; moreover, the 
fastest speeds seen in SN~2005ip persisted longer than in SN~1988Z.

(5)  We propose a model for SN~2005ip that is similar to that of 
Chugai \& Danziger (1994) for SN~1988Z, wherein a SN~II plows into a
steady but clumpy progenitor wind.  Dense clumps in the CSM persist
into the post-shock zone as they are overtaken by the forward shock
passing through the low-density region between them, while much slower
shocks are driven into individual dense clumps giving rise to the
intermediate-width components of the H$\alpha$ and He~{\sc i} lines.  
In this model, the differences between SNe~2005ip and 1988Z noted in
point (4) above can be explained if the progenitor of SN~2005ip had a
mass-loss rate a factor of $\sim$5 lower than that of SN~1988Z, making 
its CSM interaction region less massive and less optically thick.  The 
lower optical depth allows X-rays generated in the CSM interaction to
thoroughly ionize unshocked CSM gas, giving rise to the coronal spectrum.

(6) Given the rarity of coronal spectra like that of SN~2005ip, and in
light of the similarity of its inferred progenitor's wind to the
extreme conditions observed in the winds of the most luminous known
RSGs such as VY~CMa, the most straightforward interpretation seems to
be that moderate-luminosity SNe~IIn like SN~2005ip arise from extreme
RSGs having relatively high initial masses of 20--40 M$_{\odot}$.
Since the fairly modest luminosity of SN~2005ip is apparently near the
limit of what can be achieved by interaction with the densest steady
stellar winds known, this finding underscores the requirement that
exceptionally luminous SNe~IIn such as SNe 2006tf and 2006gy require
much more extreme pre-SN mass ejections analogous to massive LBV
eruptions (Smith et al.\ 2007, 2008b).  At the same time, SN~2005ip
demonstrates that not necessarily {\it all} core-collapse SNe~IIn
require this type of LBV-like mass ejection.

\acknowledgments 
\footnotesize

We acknowledge interesting discussions concerning the nature of
SN~2005ip with O.\ Fox and R.\ Chevalier.  Some of the data presented
herein were obtained at the W.M.\ Keck Observatory, which is operated
as a scientific partnership among the California Institute of
Technology, the University of California, and the National Aeronautics
and Space Administration (NASA). The Observatory was made possible by
the generous financial support of the W.M.\ Keck Foundation.  We wish
to extend special thanks to those of Hawaiian ancestry on whose sacred
mountain we are privileged to be guests. We are grateful to the staffs
at the Lick and Keck Observatories for their dedicated services. Also,
we acknowledge S.\ Park, D.\ Pooley, D.\ Poznanski, and D.S.\ Wong for
assistance with some of the observations.  KAIT was constructed and
supported by donations from Sun Microsystems, Inc., the
Hewlett-Packard Company, AutoScope Corporation, Lick Observatory, the
US National Science Foundation (NSF), the University of California,
the Sylvia \& Jim Katzman Foundation, and the TABASGO
Foundation. A.V.F.'s supernova group at U.C. Berkeley is supported by
NSF grant AST--0607485 and by the TABASGO Foundation.


\end{document}